\documentclass{article}

\usepackage[eandd,final]{neurips_2026}

\usepackage[utf8]{inputenc} 
\usepackage[T1]{fontenc}    
\usepackage{hyperref}       
\usepackage{url}            
\usepackage{booktabs}       
\usepackage{multirow}       
\usepackage{amsfonts}       
\usepackage{nicefrac}       
\usepackage{float}          
\usepackage{needspace}      
\usepackage{microtype}      
\usepackage[table]{xcolor}  
\usepackage{array}          
\usepackage{graphicx}       
\usepackage[font=small]{caption}        
\usepackage{enumitem}
\usepackage{pifont}         
\usepackage{tabularx}       
\usepackage{amsmath}        
\usepackage[toc,page]{appendix} 
\usepackage{titletoc}       
\usepackage[most]{tcolorbox} 
\usepackage{listings}       
\usepackage{xspace}

\definecolor{promptbg}{HTML}{F5F5F5}
\definecolor{promptframe}{HTML}{CCCCCC}
\newtcolorbox{promptbox}[1][]{
  breakable,
  enhanced jigsaw,
  colback=promptbg,
  colframe=promptframe,
  fonttitle=\bfseries\small,
  boxrule=0.5pt,
  arc=2pt,
  left=8pt, right=8pt, top=6pt, bottom=6pt,
  title={#1},
}

\definecolor{rustbg}{HTML}{F7F7F7}
\definecolor{rustframe}{HTML}{D0D0D0}
\definecolor{rustkw}{HTML}{8959A8}       
\definecolor{rusttrait}{HTML}{C82829}    
\definecolor{rusttype}{HTML}{4271AE}     
\definecolor{rustctor}{HTML}{1D6B4E}     
\definecolor{rustmacro}{HTML}{4271AE}    
\definecolor{ruststring}{HTML}{718C00}   
\definecolor{rustcomment}{HTML}{8E908C}  
\definecolor{rustdoccomment}{HTML}{A6791F} 
\definecolor{rustattr}{HTML}{8959A8}     

\lstdefinelanguage{rust}{
  sensitive=true,
  alsoletter={!},
  alsodigit={_},
  morekeywords={
    as, async, await, break, continue, do, dyn, else, extern, fn, for, if, in,
    impl, let, loop, match, mod, move, mut, pub, ref, return, self, Self,
    static, const, struct, super, trait, type, union, unsafe, use, where,
    while, yield, crate, box
  },
  morekeywords=[2]{
    Add, AddAssign, AsMut, AsRef, BitAnd, BitOr, BitXor, Borrow, BorrowMut,
    Clone, Copy, Debug, Default, Deref, DerefMut, Display, Div, Drop, Eq,
    Error, ExactSizeIterator, Extend, Fn, FnMut, FnOnce, From, FromIterator,
    FromStr, Hash, Hasher, Index, IndexMut, Into, IntoIterator, Iterator,
    LowerExp, LowerHex, Mul, Neg, Not, Octal, Ord, PartialEq, PartialOrd,
    Pointer, Read, Send, Shl, Shr, Sized, Sub, Sum, Sync, ToOwned, ToString,
    TryFrom, TryInto, UpperExp, UpperHex, Write
  },
  morekeywords=[3]{
    bool, char, f32, f64, i8, i16, i32, i64, i128, isize, str, u8, u16, u32,
    u64, u128, usize, unit,
    Option, Result, String, Vec, Box, Rc, Arc, Cell, RefCell, Mutex, RwLock,
    HashMap, HashSet, BTreeMap, BTreeSet, VecDeque, LinkedList, MaybeUninit,
    ManuallyDrop, NonNull, Pin, PhantomData, Cow
  },
  morekeywords=[4]{
    Some, None, Ok, Err, true, false
  },
  morekeywords=[5]{
    assert!, assert_eq!, assert_ne!, cfg!, column!, compile_error!, concat!,
    concat_idents!, dbg!, debug_assert!, debug_assert_eq!, debug_assert_ne!,
    env!, eprint!, eprintln!, file!, format!, format_args!, include!,
    include_bytes!, include_str!, line!, matches!, module_path!,
    option_env!, panic!, print!, println!, stringify!, thread_local!, todo!,
    try!, unimplemented!, unreachable!, vec!, write!, writeln!
  },
  morecomment=[l]{//},
  morecomment=[s]{/*}{*/},
  morecomment=[l][\color{rustdoccomment}\itshape]{///},
  morecomment=[l][\color{rustdoccomment}\itshape]{//!},
  morecomment=[s][\color{rustdoccomment}\itshape]{/*!}{*/},
  morestring=[b]",
  moredelim=[s][{\color{rustattr}\itshape}]{\#[}{]},
  moredelim=[s][{\color{rustattr}\itshape}]{\#![}{]},
}

\lstdefinestyle{rustcode}{
  language=rust,
  backgroundcolor=\color{rustbg},
  basicstyle=\ttfamily\footnotesize,
  keywordstyle=\color{rustkw}\bfseries,
  keywordstyle=[2]\color{rusttrait},
  keywordstyle=[3]\color{rusttype},
  keywordstyle=[4]\color{rustctor},
  keywordstyle=[5]\color{rustmacro}\itshape,
  commentstyle=\color{rustcomment}\itshape,
  stringstyle=\color{ruststring},
  showstringspaces=false,
  breaklines=true,
  breakatwhitespace=true,
  frame=single,
  rulecolor=\color{rustframe},
  framesep=4pt,
  xleftmargin=6pt,
  xrightmargin=2pt,
  aboveskip=4pt,
  belowskip=4pt,
  columns=fullflexible,
  keepspaces=true,
  literate=
    {->}{{$\rightarrow$}}2
    {=>}{{$\Rightarrow$}}2,
}

\lstnewenvironment{rustcode}{\lstset{style=rustcode}}{}

\newcolumntype{P}[1]{>{\centering\arraybackslash}p{#1}}
\newcommand{\cellszlg}{0.50cm}
\newcommand{\cellszsm}{0.40cm}
\definecolor{verylightgreennew}{RGB}{220,255,220}
\definecolor{verylightrednew}{RGB}{255,230,230}
\definecolor{verylightreddarker}{HTML}{FFCBCB}
\definecolor{lightgraynew}{rgb}{0.95,0.95,0.95}
\definecolor{darkgraynew}{rgb}{0.78,0.78,0.78}
\definecolor{greencm}{RGB}{0,153,0}
\newcommand{\cmark}{\ding{51}}
\newcommand{\xmark}{\ding{55}}
\newcommand{\cm}{{\color{greencm}\normalsize\cmark}}
\newcommand{\xm}{{\color{verylightreddarker}\normalsize\xmark}}
\newcommand\BBBBB{\rule[1.6ex]{0pt}{1.6ex}}
\newcommand{\vulnId}[1]{{\sf #1}}
\newcommand{\cellno}{\BBBBB \xm \cellcolor{verylightrednew}}
\newcommand{\cellyes}{\BBBBB \cm \cellcolor{verylightgreennew}}
\newcommand{\cellna}{\BBBBB \cellcolor{lightgraynew}}
\newcommand{\cellinvalid}{\BBBBB \cellcolor{darkgraynew}}

\definecolor{linkcolor}{HTML}{991408}  
\definecolor{citecolor}{HTML}{2E7E2A}  
\definecolor{filecolor}{HTML}{131877}  
\definecolor{menucolor}{HTML}{727500}  
\definecolor{runcolor} {HTML}{137776}  
\definecolor{urlcolor} {HTML}{0a2bbf}  
\hypersetup{colorlinks=true,linkcolor=linkcolor,citecolor=citecolor,filecolor=filecolor,menucolor=menucolor,runcolor=runcolor,urlcolor=urlcolor}
\newcommand{\sys}{\textsc{RustMizan}\xspace}

\title{\sys: A Compilable, Contamination-Aware Benchmarking Framework for Rust Vulnerabilities}

\author{%
\textbf{Tarek Elsayed}\textsuperscript{1}\thanks{Correspondence: \texttt{tareknaser360@gmail.com}}\quad
\textbf{Shiping Yang}\textsuperscript{1}\quad
\textbf{Eunsong Koh}\textsuperscript{1}\quad
\textbf{Sanika Goyal}\textsuperscript{1} \\
\textbf{Vincent Huang}\textsuperscript{1}\quad
\textbf{Paul Ngo}\textsuperscript{1}\quad
\textbf{Nathan Young}\textsuperscript{1}\quad
\textbf{Mohammad Omidvar Tehrani}\textsuperscript{1} \\
\textbf{Alvyn Kang}\textsuperscript{1}\quad
\textbf{Arnell Kang}\textsuperscript{1}\quad
\textbf{Zeyu Chen}\textsuperscript{2}\quad
\textbf{Ang\'elica Moreira}\textsuperscript{3} \\
\textbf{Xuan Feng}\textsuperscript{3}\quad
\textbf{Angel X. Chang}\textsuperscript{1,4}\quad
\textbf{Nick Sumner}\textsuperscript{1}\quad
\textbf{Steven Y. Ko}\textsuperscript{1}
\\[4pt]
\textsuperscript{1}Simon Fraser University\quad
\textsuperscript{2}Trinity University\quad
\textsuperscript{3}Microsoft Research\quad
\textsuperscript{4}Amii
}

\begin{document}

\maketitle

\begin{center}
\url{https://sfu-rsl.github.io/rust-mizan/}
\end{center}

\begin{abstract}
LLM agents are increasingly applied to vulnerability analysis, but existing
benchmarks have not kept pace. They typically rely on small non-compilable
snippets, focus on binary classification (vulnerable or not), and do not account
for the risk that publicly-released datasets are part of model training corpora.
We introduce \sys, a benchmarking framework for Rust vulnerability analysis that
addresses these gaps. \sys contains compilable code variants at the crate, file,
and function levels, with annotations for binary vulnerability detection, CWE
classification, and function- and line-level localization. A paired mutation
framework produces semantics-preserving code mutants for contamination testing
and robustness probing. Across four frontier models in an agentic setup with
command-line access, binary classification sits in the 56--65\% range and line
localization F1 is near 20\%. Adversarial cues reduce mean line F1 by about 27\%.
With low or high reasoning effort, Opus achieves about 40\% line F1 on the
95 vulnerable variants. The original models achieve 24--30\% on the same subset.
\end{abstract}

\raggedbottom
\section{Introduction}
\label{sec:intro}
\label{sec:introduction}


AI agents powered by large language models (LLMs) increasingly drive
vulnerability analysis, giving researchers and practitioners capabilities they
previously lacked. These agents analyze entire codebases as well as individual
code snippets, compile and run programs, and effectively understand and use
external tools. Beyond identifying vulnerabilities, they let users examine many
related properties, such as vulnerability classes and affected functions or
modules. Moreover, model vendors continuously train the underlying LLMs on
extensive corpora of source code from hosting platforms such as GitHub.

Existing benchmarks, however, have not kept pace with these capabilities. Many
recent benchmarks for AI agents target general software
development~\citep{2024_swebench, 2026_terminalbench, zan2026multiswebench},
which differs fundamentally from vulnerability analysis. On the
other hand, existing vulnerability benchmarks often rely on small code snippets
drawn from version-control diffs, stripping away the build environment necessary
to compile and execute the code. They typically focus on binary detection
(vulnerable or not), neglecting related properties such as vulnerability class
and relevant contextual information. This setup cannot exercise the capabilities
thorough analysis requires, e.g., using tools such as static analyzers and
fuzzers, interacting with real build systems and codebases, and iterating on the
analysis~\citep{2025_vuln_survey, 2025_primevul}. They also suffer from data
contamination: once a dataset is released, LLMs are trained on it, and the
models then recall the vulnerabilities from memory instead of using their
reasoning abilities~\citep{2026_benchmark_survey, 2025_primevul, 2023_vjbench}.

To address these limitations, we introduce \sys (mizan in Arabic means
\emph{scale} or \emph{balance}), a benchmarking framework for Rust vulnerability
analysis. We focus on Rust as security-critical infrastructure increasingly
adopts it, including the Linux kernel, Android, and Windows, yet it remains
underrepresented in vulnerability benchmarks~\citep{2026_benchmark_survey}.
Using Rust as a concrete example, \sys showcases the following three design
principles that, to the best of our knowledge, no other vulnerability benchmarks
provide in combination, including Rust
benchmarks~\citep{2020_rust_memory_bugs, 2021_rust_cves,
2023_rust_security_risks, androutsopoulos2025deepsurfdetectingmemorysafety,
10.1145/3728890, xiang2026rustsweBench, khatry2025crustbench}:
\begin{enumerate}[leftmargin=*, itemsep=0pt, topsep=0pt]

  \item \textbf{Multi-task evaluation:} \sys supports four complementary
  tasks---binary vulnerability classification, CWE classification,
  function-level localization, and line-level localization, covering the
  pipeline from detection to precise diagnosis.
  
  \item \textbf{Multi-level compilable variants:} For each vulnerability, \sys
  constructs a family of compilable program \emph{variants} at up to three context
  levels: \emph{crate-level} (entire Rust package), \emph{file-level} (the file
  containing the vulnerability and required dependencies), and
  \emph{function-level} (the vulnerable function and required dependencies).
  Each available variant is a standalone, compilable program that preserves
  the same vulnerability.
  Compilability enables evaluation with traditional program analysis tools
  (e.g., static analyzers, fuzzers) and, crucially, with LLM agents that can
  compile, explore, and interact with the code. Multi-level variants support
  controlled studies of how context granularity affects analysis accuracy.
  
  
  \item \textbf{Built-in support for contamination mitigation and robustness
  testing:} \sys provides automated semantics-preserving code mutants that test
  whether models rely on memorization rather than code reasoning. \sys also
  provides code mutants that test how models behave against adversarial
  cues such as misleading comments.
  
\end{enumerate}


\sys includes a dataset of 42 real-world memory-safety Common Vulnerabilities and Exposures (CVEs), yielding 173
compilable code variants at the crate, file, and function levels. We pair each
CVE with ground-truth annotations for all four tasks (binary classification, Common Weakness Enumeration (CWE) classification, function localization, and line localization), and the mutation
framework automatically generates contamination-testing and adversarial
variants. Although \sys focuses on Rust, we believe that its design
principles---multi-task evaluation, multi-level compilable variants, and
built-in support for contamination mitigation and robustness testing---extend
naturally to other programming languages.

We evaluate four frontier models (Claude Sonnet~4.6, GPT~5.4, Gemini~3.1~Pro,
and Qwen~3.6~Plus) and address four research questions:
\begin{enumerate}[leftmargin=*, itemsep=0pt, topsep=0pt, label=\textbf{RQ\arabic*.}]
  \item \textbf{How do frontier models perform across the full vulnerability
  analysis pipeline?} The four models reach 56--65\% detection accuracy, but line-level
  localization F1 stays around 20\%, well below detection
  (Section~\ref{sec:main_results}).
  \item \textbf{How does code context granularity affect analysis accuracy?}
  Function-level context (the most narrow context) substantially improves
  vulnerability localization (Section~\ref{sec:context_level}).
  \item \textbf{Do models rely on memorized vulnerability patterns?}
  Under the tested mutations and settings, aggregate Line F1 shows no clear
  contamination signal for the four frontier models, though trajectory analysis
  reveals one model explicitly recalling CVE identifiers. Open-source models
  fine-tuned on our vanilla dataset exhibit Line F1 inflation that Benign and
  Malignant mutations substantially reduce
  (Section~\ref{sec:contamination_results}).
  \item \textbf{Are models robust to adversarial cues?} Misleading comments and
  identifier renames cause about a 27\% relative drop in mean line localization F1,
  showing that models follow these cues rather than grounding analysis in code
  semantics (Section~\ref{sec:robustness_results}).
\end{enumerate}

We envision \sys as a platform for accelerating the development of better LLM
agents for vulnerability analysis and for studying the effect of context
granularity and contamination. We release the complete framework, dataset,
evaluation scripts, and all agent trajectories, along with a public
leaderboard. Licensing and Responsible AI documentation are in
Appendix~\ref{sec:artifact_card}.

\section{\sys}
\label{sec:framework}

\begin{figure}[t]
  \centering
  \includegraphics[width=\linewidth]{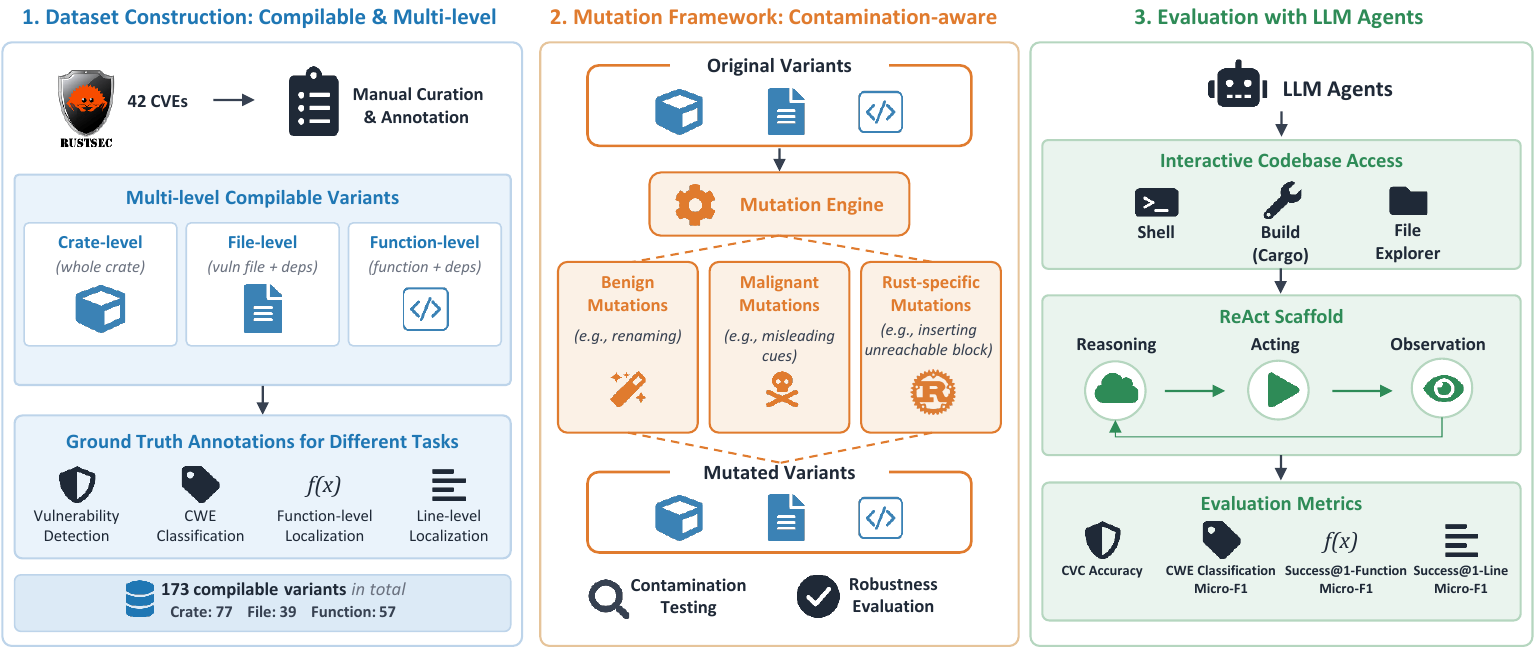}
  \caption{Overview of \sys. (1) a manually curated dataset of 42 RustSec CVEs packaged as 173 multi-level compilable variants with task-specific ground truth, (2) a mutation framework that produces benign, malignant, and Rust-specific code mutants for contamination testing and robustness evaluation, and (3) an LLM-agent evaluation pipeline using a ReAct scaffold with interactive codebase access, scored across four vulnerability-analysis tasks.
  } \label{fig:usage_model}
  \vspace{-1em}
  \end{figure}

\sys has two components. The first is a dataset of real-world Rust memory-safety
CVEs, packaged as multi-level compilable variants with peer-reviewed annotations
(Section~\ref{sec:dataset}). The second is an extensible mutation framework that
we pair with the dataset; it produces semantics-preserving mutated datasets for
contamination mitigation and adversarial robustness testing
(Section~\ref{sec:mutation_framework}). Figure~\ref{fig:usage_model} shows how
the two components fit together. Table~\ref{tab:benchmark_comparison_small}
compares \sys to other vulnerability and Rust benchmarks, showing that \sys is
the only compilable, multi-level, contamination-aware benchmarking framework, to
the best of our knowledge. Appendix~\ref{sec:benchmark_comparison} compares \sys
to other benchmarks more comprehensively.


\definecolor{mizanrow}{gray}{0.92}
\definecolor{familyrow}{gray}{0.97}

\begin{table}[H]
\centering
\scriptsize
\setlength{\tabcolsep}{4pt}
\renewcommand{\arraystretch}{1.18}
\caption{RustMizan in context of two families of related benchmarks. \checkmark{}\,=\,all entries support it; $\sim$\,=\,some do; \ding{55}\,=\,none. \textbf{Vuln.}: full vulnerability-analysis pipeline (detection, CWE classification, function-/line-level localization). \textbf{Comp.}: code compiles. \textbf{M-lvl}: same task at multiple code granularities. \textbf{Cont.}: built-in contamination handling. \textbf{Rust}: targets Rust. RustMizan is the only family that satisfies all five.}
\label{tab:benchmark_comparison_small}
\begin{tabularx}{\linewidth}{@{}>{\raggedright\arraybackslash}X >{\raggedright\arraybackslash}p{3.0cm} c c c c c@{}}
\toprule
\textbf{Benchmarks}
& \textbf{Task}
& \textbf{Vuln.}
& \textbf{Comp.}
& \textbf{M-lvl}
& \textbf{Cont.}
& \textbf{Rust}
\\

\midrule
\rowcolor{familyrow}
\multicolumn{7}{@{}l}{\textsc{\textbf{Vulnerability benchmarks}}\,$^{a}$} \\
BigVul, CVEfixes, CrossVul, Devign, DiverseVul, PrimeVul, CVE-Bench, CyberGym, SEC-bench, JITVul
& Detection, repair, exploit, PoC reproduction
& $\sim$ & $\sim$ & \ding{55} & $\sim$ & \ding{55}
\\

\midrule
\rowcolor{familyrow}
\multicolumn{7}{@{}l}{\textsc{\textbf{Rust datasets / benchmarks}}\,$^{b}$} \\
Memory-safety in Rust, All Rust CVEs, Rust ecosystem, Rust-SWE-bench, deepSURF, Safe4U, CRUST-Bench
& Study, issue resolution, detection, transpilation
& $\sim$ & $\sim$ & \ding{55} & \ding{55} & \checkmark
\\

\midrule
\rowcolor{mizanrow}
\textbf{RustMizan (ours)}
& Vulnerability analysis (detection, CWE, fn-loc, line-loc)
& \textbf{\checkmark} & \textbf{\checkmark} & \textbf{\checkmark} & \textbf{\checkmark} & \textbf{\checkmark}
\\
\bottomrule
\end{tabularx}
\\[0.4em]
\begin{flushleft}
\scriptsize
$^{a}$\,\citep{10.1145/3379597.3387501,10.1145/3475960.3475985,10.1145/3468264.3473122,10.5555/3454287.3455202,10.1145/3607199.3607242,2025_primevul,2025_cvebench,zhu2025cvebench,wang2026cybergym,lee2025secbench,yildiz-etal-2025-benchmarking}.
$^{b}$\,\citep{2020_rust_memory_bugs,2021_rust_cves,2023_rust_security_risks,xiang2026rustsweBench,androutsopoulos2025deepsurfdetectingmemorysafety,10.1145/3728890,khatry2025crustbench}.
\end{flushleft}
\vspace{-1em}
\end{table}

We expect researchers \emph{not} to use our dataset as-is. Instead, we expect
them to apply mutations to the vanilla dataset and use the mutated dataset to
evaluate their vulnerability analysis techniques. This reduces the risk of LLMs
recalling our vulnerabilities from memory after our dataset is released and
potentially seen during training. It also allows researchers to conduct
robustness testing (Section~\ref{sec:mutation_framework}).

\subsection{Dataset}
\label{sec:dataset}

\sys focuses on Rust memory-safety vulnerabilities. Despite Rust's guarantees against common memory bugs, prior studies show that Rust programs still suffer from critical vulnerabilities, with memory and concurrency problems making up almost two-thirds of reported issues~\citep{2020_rust_memory_bugs,2020_rust_unsafe_usage,2020_xrust,2023_rust_security_risks}. Due to the language's relative youth, only a few early datasets exist for studying Rust vulnerability patterns~\citep{2020_rust_memory_bugs,2021_rust_cves,2023_rust_security_risks}. At the same time, Rust is increasingly adopted in safety-critical systems including the Linux kernel, Android, and Windows, yet it remains under-represented in vulnerability benchmarks~\citep{2026_benchmark_survey}.


%
%
%
%
%
%
%

\begin{figure}[t]
  \centering
  \includegraphics[width=\linewidth]{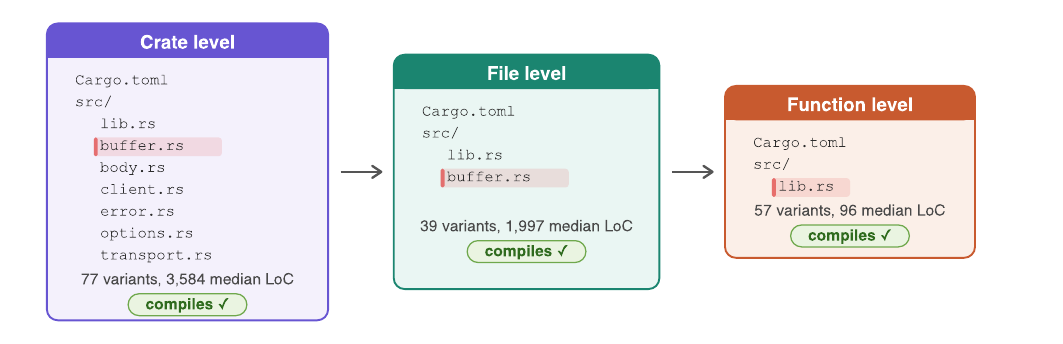}
  \caption{Example of a CVE in \sys with three standalone compilable variants of decreasing scope: the full crate, a single-file reduction with its compilation dependencies, and a single-function reduction. Red marks the vulnerable file, tracked across all three levels.}
  \label{fig:multi_level_variants}
  \vspace{-1em}
\end{figure}

\textbf{Key Property:} The most important property of our dataset is that it
consists of \emph{multi-level compilable program variants}. Every variant in
\sys is a standalone compilable program that represents one of the three context
levels: full crate (Rust package for an entire program), single file, and
individual function. A full crate variant is an entire program that contains a
vulnerability. A file-level variant is stripped down to the file containing the
vulnerability plus any other files and type definitions needed to compile it,
packaged as a standalone crate. A function-level variant is further reduced to
just the vulnerable function and its compilation dependencies. If a
vulnerability spans multiple files or functions, we include all of them.
Not every vulnerability has all three context levels. For example, we skip the
function-level variant when the file is essentially a single function.
Figure~\ref{fig:multi_level_variants} illustrates the reduction for a single
CVE.

For each vulnerability, we construct up to six variants: three (crate, file,
function) from the vulnerable version and three from the patched version (when a
fix is available).
We construct all variants manually and verify that they compile. While
labor-intensive, this ensures high data quality: most large vulnerability
datasets rely on automated labeling, which introduces significant label
noise~\citep{2025_primevul}.

These multi-level, compilable variants serve three purposes. First, they enable
evaluation with LLM-based agents that can compile the code, inspect files, trace
function calls, and use any tools available through a shell. Previous
vulnerability benchmarks that only provide non-compilable snippets make such
evaluation impossible. Second, they provide an avenue to study how the amount of
context (i.e., other code present) affects analysis accuracy. Since it is the
same vulnerability at different levels, any performance difference is due to the
amount of context. Third, they enable the evaluation of traditional (non-LLM)
program analysis tools that require compilation, such as static analyzers that
operate on compiler intermediate representations. We demonstrate this with
Kani~\citep{2024_kani} and RAPx~\citep{2025_rapx} in
Appendix~\ref{sec:traditional_tools}.


\textbf{Dataset Curation:} Each vulnerability in our dataset undergoes a
rigorous curation process where two reviewers (primary and secondary)
independently verify the vulnerability. The primary reviewer first selects a
vulnerability from the RustSec Advisory Database~\citep{2025_rustsec}, a
community-maintained repository of security advisories for Rust crates. We use
RustSec due to its recognized authority in the Rust community, though our design
is not restricted to a particular source. The primary reviewer annotates the
chosen vulnerability with its CVE identifiers, CWE classifications, year of
disclosure, and precise locations of the vulnerability, i.e., functions and
lines. These annotations come from CVE descriptions, GitHub issue discussions,
commit messages, and code review. Once that is done, the secondary reviewer
verifies the information by independently repeating the same steps. If there are
differences, the two reviewers iteratively resolve them until they reach a
consensus. The final annotations support four evaluation tasks: binary
vulnerability classification (classifying if a program contains a vulnerability
or not), CWE classification, function-level localization, and line-level
localization.



\textbf{Dataset Statistics:} The dataset covers 42 memory-safety CVEs (35 with
patches, 83\%) yielding 173 compilable variants in total (77 crate-, 39 file-,
57 function-level), disclosed between 2018 and 2025. Per-level median sizes are
shown in Figure~\ref{fig:multi_level_variants}. While the total number of CVEs
is modest, we manually curate each vulnerability and construct compilable,
multi-level variants. Thus, our dataset prioritizes quality and control over raw
count. Nevertheless, our dataset is easily extensible and we are actively
adding more vulnerabilities.

\subsection{Mutation Framework}
\label{sec:mutation_framework}

We pair \sys with an extensible mutation framework for contamination mitigation
and robustness testing. Prior work has shown that semantics-preserving code
mutations are effective for detecting memorization:
CodeMorph~\citep{2025_codemorph} found average accuracy drops of 25\% on code
completion tasks, and DEFECTS4J-TRANS~\citep{2025_defects4j_trans} reported up
to 49\% reduction in correct patches after mutation. Other work has explored
semantic-preserving mutations that add misleading safety cues, such as comments
(e.g., ``// The next line is vulnerable.'') and safety-implying renames, to
confuse LLMs and reduce vulnerability detection
accuracy~\citep{2024_secllmholmes}. Building on these insights, we design \sys
to support mutations for contamination mitigation and robustness evaluation. As
mentioned earlier, we expect researchers to apply mutations to our vanilla
dataset and use the mutated dataset to evaluate their vulnerability analysis
techniques.



\textbf{Mutation categories:} Mutations fall into three categories.
\emph{Benign} mutations apply semantics-preserving mutations, such as
formatting changes, for-to-while loop rewrites, and variable or function
renaming, to reduce the risk of contamination. \emph{Malignant} mutations
inject adversarial safety cues, such as misleading comments or variable names,
to evaluate whether agents are robust to misleading signals. The first two
categories rely only on common language features (comments, loops, identifiers)
and apply naturally to languages beyond Rust. \emph{Rust-Specific} mutations
leverage Rust features, such as reordering \texttt{\#[derive(..)]}
attributes. Other language-specific work has introduced similar
tailored mutations, e.g., C-focused work adds safe uses of functions like
\texttt{strcpy} as non-trivial mutations~\citep{2024_secllmholmes}, and
Java-focused work adds Java-specific structural edits~\citep{2023_vjbench}. The
full specification of all implemented mutations is provided in
Appendix~\ref{sec:mutation_catalog}.

\textbf{Implementation:} Applying these mutations to compilable vulnerability
benchmarks poses challenges that prior work has not addressed
(Table~\ref{tab:contamination_comparison}). Some prior
approaches~\citep{2023_vjbench, 2023_recode} apply mutations to isolated code
snippets without verifying compilability afterward, which is simpler than
mutating full projects where, for example, renaming a symbol requires updating
every reference across files. Others~\citep{2025_codemorph,
2025_defects4j_trans} use LLMs to apply mutations and include compilation
checks, but rely on human annotators to verify semantic equivalence. Crucially,
none target vulnerability analysis, so they need not preserve any ground truth
about the vulnerability. Applying mutations to vulnerability benchmarks is
fundamentally harder, because mutations can change the ground truth. For
example, inserting dead code shifts line numbers and breaks line-level labels,
and renaming a function invalidates annotations that reference it by name.

\sys addresses these challenges primarily by using tools from the Rust compiler
ecosystem. Formatting mutations are applied via \texttt{rustfmt}, the official
Rust formatter. AST-based structural mutations use the \texttt{syn} and
\texttt{quote} crates, which provide Rust's standard AST parsing and code
generation. Symbol renaming uses \texttt{rust-analyzer}, the official Rust
language server, which correctly resolves and updates all references across the
entire project. The framework rejects transformations that fail compilation.
Regression testing provides empirical evidence of behavior preservation for
successfully applied mutations. We report validation results for 17 mutation
operators across four open-source Rust crates in
Appendix~\ref{subsec:mutation_preservation}.

For ground truth maintenance, \sys uses three mechanisms. For positional shifts
from formatting and code-insertion mutations, we use \emph{marker tracking}. We
insert a unique comment marker before each vulnerable line, move the marker with
the vulnerable line through the mutation, and track them together. For AST-based
mutations, where comments are stripped during parsing, we use
\emph{content-based tracking}. The framework matches the exact text of each
vulnerable line before and after the mutation. For symbol-renaming mutations, we
use \emph{rename validation}. We keep a per-mutation map of renamed symbols. In
addition, our mutations are conservative, i.e., we only mutate clear-cut cases
that do not have any ambiguity. Together, these ensure that function-level and
line-level annotations remain
accurate after a combination of mutations are applied. The framework applies
mutations sequentially to individual variants and automatically rolls back to
the backed-up original if any step fails.
Appendix~\ref{sec:ground_truth_tracking} describes the full mutation pipeline
(Figure~\ref{fig:mutation_framework}) along with the three tracking mechanisms
in more detail and the per-variant log of mutation outcomes.



\begin{table}[!hb]
\centering
\small
\caption{Comparison of mutation-based contamination mitigation approaches. \checkmark~= full support, ${\sim}$~= partial, \ding{55}~= none.
\emph{ReCode} \& \emph{DEFECTS4J-TRANS}: partial LLM use for variable renaming; no human involvement in applying mutations but annotator checks are performed. \emph{ReCode:} targets snippets in code generation benchmarks and does not address cross-file support. \emph{CodeMorph}: fully LLM-driven (mutations selected and applied by LLMs), partial human dependence via annotator checks for semantic equivalence; guarantees cross-file support and compilability, though uncompilable variants (due to LLM hallucinations) are discarded.
}
\label{tab:contamination_comparison}
\resizebox{\textwidth}{!}
{
\begin{tabular}{@{}lccccl@{}}
\toprule
\textbf{Approach} & \textbf{No LLM} & \textbf{No Human} & \textbf{Cross-file} & \textbf{GT} & \textbf{Target} \\
& \textbf{Required} & \textbf{in Loop} & \textbf{Support} & \textbf{Tracking} & \textbf{Tasks} \\
\midrule
VJBench-trans \citep{2023_vjbench} & \checkmark & \ding{55} & \checkmark & \ding{55} & Vuln.\ Fixing \\
ReCode \citep{2023_recode} & $\sim$ & $\sim$ & \ding{55} & \ding{55} & Code Gen. \\
CodeMorph \citep{2025_codemorph} & \ding{55} & $\sim$ & \checkmark & \ding{55} & Code Compl. \\
DEFECTS4J-TRANS \citep{2025_defects4j_trans} & $\sim$ & $\sim$ & \checkmark & \ding{55} & Program Repair \\
\midrule
\textbf{\sys (Ours)} & \checkmark & \checkmark & \checkmark & \checkmark & Vuln.\ Analysis \\
\bottomrule
\end{tabular}
}
\vspace{-1em}
\end{table}

\section{Evaluation}
\label{sec:evaluation}

Our evaluation shows how LLM agents perform on our dataset and mutation
framework, addressing the four research questions introduced in
Section~\ref{sec:intro}.

\subsection{Setup}
\label{sec:eval_setup}
\label{sec:agentic_harness}

\textbf{Agent and model setup:} We run each LLM agent in a sandboxed Docker
environment, built with Inspect-AI~\citep{2024_inspectai}. We evaluate Claude
Sonnet~4.6, GPT~5.4, Gemini~3.1~Pro, and Qwen~3.6~Plus, each driven by a ReAct
(Reasoning + Acting) scaffold~\citep{2023_react} that alternates brief reasoning
steps with targeted tool calls and feeds the resulting observations into
subsequent decisions. The agent has shell access and analyzes one code variant
at a time. The agent can freely explore the codebase, compile the code, and
read any file before producing its analysis. This setup reflects how interactive
vulnerability analysis works in practice: the model decides what to examine and
in what order, rather than receiving a pre-selected snippet. The task prompt
asks the agent to produce a structured JSON response covering the four
evaluation tasks: whether the code is vulnerable, what CWE applies, which
functions are vulnerable, and which lines are vulnerable. We log all agent
steps and reasoning traces, enabling trajectory analysis that can reveal
contamination signals such as models recalling CVE identifiers from memory
(Section~\ref{sec:contamination_results}).
We use the same ReAct prompt for all four models, with a fixed JSON output
format that includes an \texttt{explanation} field to encourage reasoning
before classification, so reported performance differences reflect model
capability rather than prompt-engineering effort. The full task description,
agent system prompt, and how we have chosen our prompting strategy are provided
in Appendix~\ref{sec:evaluation_prompts}. The vanilla evaluation consumed 24.5M
tokens for Claude Sonnet~4.6, 20.0M for GPT~5.4, 47.7M for Gemini~3.1~Pro, and
150.8M for Qwen~3.6~Plus. Model configurations are reported in
Table~\ref{tab:model_configuration}.

\textbf{Metrics:} For each code variant, we compare the model's structured JSON
response against the ground truth across four tasks. \emph{CVC Accuracy} (Crate
Vulnerability Classification) scores 1 when the predicted vulnerable or patched
status matches ground truth (and 0 otherwise), averaged over all 173 variants,
including the 78 patched variants. \emph{Success@1-Function} and
\emph{Success@1-Line} score 1 when a single agent run identifies at least
one ground-truth vulnerable function or line, respectively (and 0 otherwise).
Both rates are averaged over the 95 vulnerable variants. CWE classification, function
localization, and line localization are set-based metrics: we compute TP, FP, FN
by set intersection between model answers and ground truth, and report
micro-averaged F1 scores. Full per-metric definitions and formulas are given in
Appendix~\ref{sec:evaluation_metrics}.

\textbf{Dataset mutants:} We evaluate on four dataset categories constructed
using our mutation framework, as described in
Section~\ref{sec:mutation_framework}.
Namely, \emph{Vanilla} uses the original unmutated code. \emph{Benign} applies
semantic-preserving, neutral mutations, such as loop rewrites. It tests whether
models rely on memorized patterns. \emph{Malignant} injects adversarial safety
cues, such as misleading comments. It tests robustness to code-level deception.
\emph{Rust-Specific} applies semantic-preserving edits only possible in Rust.
It tests the effect of language-specific mutations.

\subsection{Models Detect but Struggle to Localize (RQ1)}
\label{sec:main_results}

\definecolor{zebrarow}{gray}{0.93}
\newcommand{\best}[1]{\textbf{#1}}

\begin{figure}[t]
\centering
\includegraphics[width=.85\linewidth]{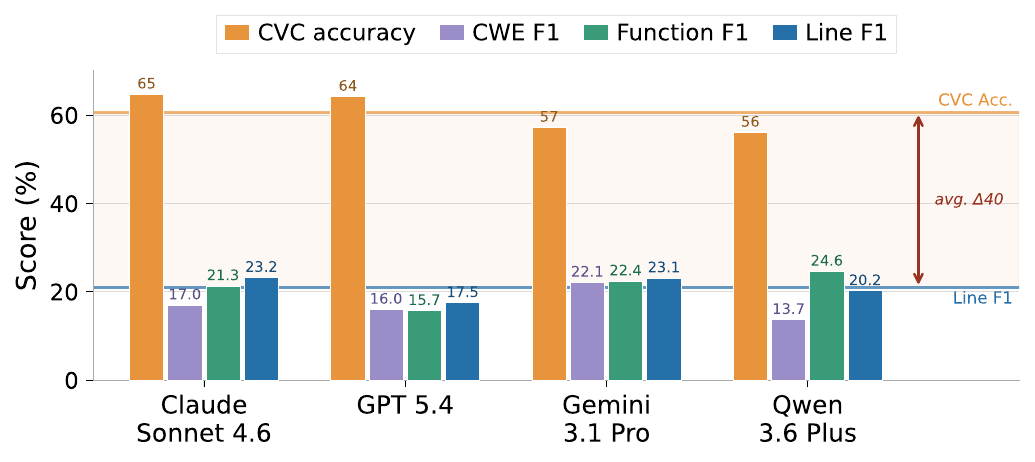}
\caption{Detection and localization scores on \sys (vanilla). Detection accuracy is higher than localization F1 for all four models. The shaded band marks the gap between average detection accuracy and average line localization F1.
}
\label{fig:detection_localization_gap}
\vspace{-1em}
\end{figure}

Table~\ref{tab:main_results} shows performance on the vanilla dataset, and
Figure~\ref{fig:detection_localization_gap} plots the same numbers side by side.
Always predicting ``vulnerable'' gives 54.9\% CVC Accuracy (95/173).
For these four models, the main finding is the gap between detection and localization. CVC Accuracy
sits in the 56--65\% range, but Line F1 only reaches 17--23\%. A benchmark that
tested detection alone would make these models look much more capable than they
actually are at pinpointing where the issue is. This gap motivates the
multi-task design.

Among these four models, no single model clearly outperforms the others across all tasks: Claude 4.6 leads on detection and line F1, Gemini on CWE F1 and function Success@1, and Qwen on function F1. Model ranking depends on which task is measured, which is the kind of comparison the multi-task design is intended to support. CWE classification is also weak across the board (F1 from 13.7 to 22.1) despite models having full access to the code. A per-variant breakdown of line localization across all four models is in Appendix~\ref{sec:per_sample_results}.

We also evaluate different reasoning effort levels
(Appendix~\ref{sec:reasoning_effort}).

\begin{table}[!hb]
\centering
\caption{Vulnerability analysis results on \sys (vanilla). CVC Accuracy and Success@1 are binary metrics with raw counts shown; F1 scores are micro-averaged. Highest point estimates per column in \textbf{bold}. 95\% bootstrap intervals are in Appendix~\ref{sec:uncertainty_estimates}.}
\label{tab:main_results}
\small
\begin{tabular}{@{}l cc cc cc@{}}
\toprule
& \textbf{Detection} & \textbf{Classification} & \multicolumn{2}{c}{\textbf{Localization (F1)}} & \multicolumn{2}{c}{\textbf{Success@1}} \\
\cmidrule(lr){2-2} \cmidrule(lr){3-3} \cmidrule(lr){4-5} \cmidrule(lr){6-7}
\textbf{Model} & \textbf{CVC Acc.} & \textbf{CWE F1} & \textbf{Func.} & \textbf{Line} & \textbf{Func.} & \textbf{Line} \\
\midrule
\rowcolor{zebrarow}
Claude Sonnet 4.6 & \best{64.7} {\scriptsize(112/173)} & 17.0 & 21.3 & \best{23.2} & 46.3 {\scriptsize(44/95)} & \best{47.4} {\scriptsize(45/95)} \\
GPT 5.4           & 64.2 {\scriptsize(111/173)}        & 16.0 & 15.7 & 17.5        & 46.3 {\scriptsize(44/95)} & 46.3 {\scriptsize(44/95)} \\
\rowcolor{zebrarow}
Gemini 3.1 Pro    & 57.2 {\scriptsize(99/173)}         & \best{22.1} & 22.4 & 23.1 & \best{52.6} {\scriptsize(50/95)} & 36.8 {\scriptsize(35/95)} \\
Qwen 3.6 Plus     & 56.1 {\scriptsize(97/173)}         & 13.7 & \best{24.6} & 20.2 & 44.2 {\scriptsize(42/95)} & 43.2 {\scriptsize(41/95)} \\
\bottomrule
\end{tabular}
\vspace{-1em}
\end{table}

\begin{figure}[H]
\centering
\includegraphics[width=.9\textwidth]{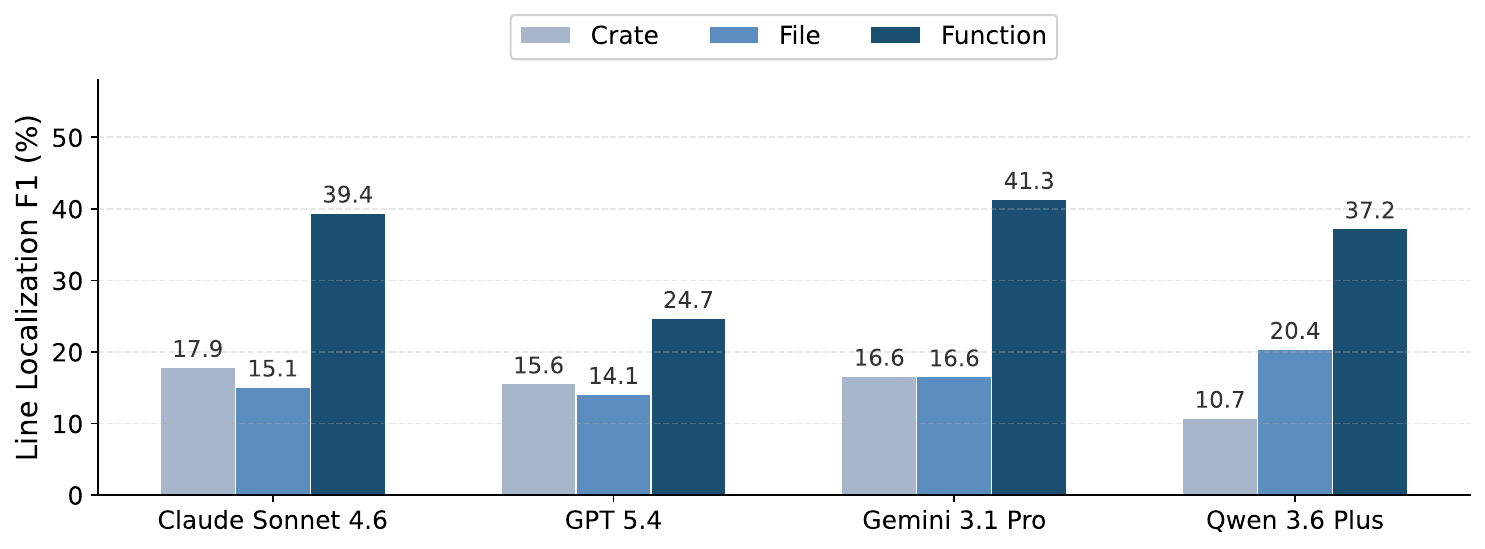}
\vspace{-1em}
\caption{Line localization F1 across context levels (crate, file, function). Function-level context consistently helps. The crate-to-file differences vary across the models.
}
\label{fig:performance_by_level}
\vspace{-0.5em}
\end{figure}

\subsection{Narrow Context Improves Localization (RQ2)}
\label{sec:context_level}

Figure~\ref{fig:performance_by_level} shows Line F1 across the three context
levels. Function-level context yields the highest scores for all four models:
Claude moves from 17.9 at crate level to 39.4 at function level, GPT from 15.6
to 24.7, Gemini from 16.6 to 41.3, and Qwen from 10.7 to 37.2. This kind of
comparison is enabled by the multi-level design, although the available
vulnerabilities differ across levels.
Narrower context also reduces the search space, which can contribute to these
localization gains.


\subsection{No Clear Contamination Signal (RQ3)}
\label{sec:contamination_results}

We run all four models on the dataset mutants described in
Section~\ref{sec:eval_setup} (Table~\ref{tab:mutation_results}). Benign
mutations change Line F1 by only 0.5 points on average across the four models,
and Rust-Specific mutations sit at a similar 0.7 point change. The 95\% intervals
for both mean changes include zero (Appendix~\ref{sec:uncertainty_estimates}). Prior work has
interpreted significant performance degradation under semantic-preserving
mutations as evidence of LLM contamination, with reported drops of 24--49\% on
related code tasks~\citep{2023_recode, 2025_codemorph, 2025_defects4j_trans,
2023_vjbench}. We observe no comparable drop in Line F1, giving no clear
contamination signal under the tested mutations and settings.

\begin{table}[!hb]
\centering
\caption{Line localization micro-F1 across dataset mutants on \sys. The last row reports the mean change relative to Vanilla across the four models. Bootstrap intervals for scores and paired changes are in Appendix~\ref{sec:uncertainty_estimates}.}
\label{tab:mutation_results}
\smallskip
\small
\begin{tabular}{@{}lcccc@{}}
\toprule
\textbf{Model} & \textbf{Vanilla} & \textbf{Benign} & \textbf{Rust-Specific} & \textbf{Malignant} \\ 
\midrule
Claude Sonnet 4.6 & 23.2 & 19.7 & 24.0 & \textbf{12.8} \\
GPT 5.4           & 17.5 & 16.4 & 15.5 & 14.3 \\
Gemini 3.1 Pro    & 23.1 & 20.7 & 22.2 & 19.7 \\
Qwen 3.6 Plus     & 20.2 & 25.2 & 19.6 & 14.7 \\
\midrule
\textbf{Mean $\Delta$ vs Vanilla} & -- & $-0.5$ & $-0.7$ & $\mathbf{-5.6}$ \\
\bottomrule
\vspace{-1em}
\end{tabular}

\end{table}

To assess the mutation framework under known contamination, we
have conducted a controlled contamination experiment. We have fine-tuned three
open-source models on the Vanilla dataset, and repeated the same experiment. Our
results show that fine-tuned models report higher Line F1 by
47--72 points, but both Benign and Malignant mutations suppress Line F1 by
44--77 points, showing that these mutations reduce contamination-associated
line-localization score inflation in the tested models. The details are shown in
Appendix \ref{sec:data_decontamination}.


Beyond aggregate metrics, the evaluation setup logs full model trajectories for
inspection. To illustrate the value of these logs, we apply
Docent~\citep{2025_docent}, a system for summarizing, clustering, and searching
agent transcripts, to a small subset of trajectories and find one case where
Gemini 3.1 Pro explicitly recalls a CVE identifier and references specific line
numbers from memory, despite the prompt containing no advisory information. We
note that chain-of-thought faithfulness is an active area of research and
reasoning traces may not always reflect a model's true internal
process~\citep{2025_cot_faithfulness}. An example trace is reproduced in
Appendix~\ref{sec:trajectory_examples} (Figure~\ref{fig:gemini_docent}).

\subsection{Adversarial Cues Degrade Localization (RQ4)}
\label{sec:robustness_results}

Malignant mutations lower the point estimates for all four models
(Table~\ref{tab:mutation_results}). Line localization F1 drops by 5.6 points on
average (95\% CI: 2.3--9.9 points), about a 27\% relative decrease from the Vanilla mean of 21.0. The
largest drop is on Claude (23.2 to 12.8). Qwen falls by 5.5 points, Gemini by
3.3, and GPT by 3.2. The per-model change intervals exclude zero for Claude and
Qwen, but include zero for GPT and Gemini. The degradation indicates that the agents are susceptible
to adversarial cues such as misleading comments and identifier renames. Since
adversaries can inject such adversarial cues into a codebase, it is important to
evaluate robustness to such tactics for agents deployed in production.

\section{Related Work}
\label{sec:related}

\textbf{Vulnerability detection datasets:}
Recent surveys have converged on a consistent set of problems with existing vulnerability benchmarks.
Most work restricts itself to binary classification of function-level vulnerabilities, with insufficient attention to multi-file dependencies, and C/C++ as the dominant language~\citep{2025_vuln_survey}.
Automatic labeling, used by most large datasets because manual labeling is expensive, introduces significant label noise, and current evaluation metrics fail to capture practical utility~\citep{2025_primevul}.
A meta-analysis of 274 code-related benchmarks finds that nearly 80\% do not handle data contamination, quality checks are often neglected, and language coverage is heavily skewed: dozens of languages have at most one benchmark~\citep{2026_benchmark_survey}.
For Rust specifically, prior work falls into two categories: empirical studies of CVEs and unsafe-code patterns, and narrow-scope detection tools for specific bug classes~\citep{2020_rust_memory_bugs, 2021_rust_cves, 2023_rust_security_risks, androutsopoulos2025deepsurfdetectingmemorysafety, 10.1145/3728890}; and more recently, benchmarks that target other problems entirely such as agentic issue resolution and C-to-Rust transpilation~\citep{xiang2026rustsweBench, khatry2025crustbench}.
\sys addresses these gaps: it evaluates four tasks including binary detection, provides manually labeled and peer-reviewed annotations, includes built-in contamination mitigation, and targets Rust, a safety-critical but under-represented language.

\textbf{Agentic benchmarks:}
There has been considerable effort to benchmark LLM agents across domains, including SWE-bench for software engineering~\citep{2024_swebench}, WebArena for web interaction~\citep{2024_webarena}, and Terminal-Bench for command-line tasks~\citep{2026_terminalbench}.
Within security, the closest work to ours is CVE-Bench, which provides 509 CVEs across four languages and gives agents a test environment for generating vulnerability repairs~\citep{2025_cvebench}.
However, CVE-Bench targets vulnerability \emph{repair} rather than \emph{analysis}, does not address training data contamination, and does not provide code at multiple context levels for studying the tradeoff between input scope and analysis accuracy.
\sys targets the step that precedes repair: given a codebase, determine whether vulnerabilities exist, classify their types, and localize them.

\textbf{Data contamination in code benchmarks:}
Data contamination, where evaluation data overlaps with training corpora, is an active area of research.
Techniques have been proposed to both detect contamination~\citep{2020_gpt3,2024_minkprob,2024_contamination_detection,2018_privacy_overfitting,2024_memorization_code} and mitigate it, through approaches such as temporal freshness~\citep{2023_stop_uploading,2024_livecodebench,2024_condefects}, access restriction, and semantic-preserving code mutations~\citep{2023_vjbench,2023_recode,2025_codemorph,2025_defects4j_trans}.
Mutation-based methods have been applied to code generation, completion, vulnerability fixing, and program repair, typically revealing performance drops of 25--49\% that expose memorization~\citep{2023_vjbench,2023_recode,2025_codemorph,2025_defects4j_trans,2026_euraste}.
\sys is designed to directly support this line of research: it integrates established mutation operations into a benchmarking framework where researchers can easily apply and extend them.

\textbf{Rust program analysis tools:} Program analyzers at different levels
(e.g. Miri, MIR, binary) exist for
Rust~\citep{miri,2024_kani,2025_rapx,10.1145/3477132.3483570,10.1145/3460120.3484541,10.1145/3542948,10.1145/3642974.3652281,mirai},
all of which require compilable code. \sys's compilable multi-level variants are
directly usable by such tools; we survey them in
Appendix~\ref{sec:rust_program_analysis_tools} and demonstrate two of them on
\sys in Appendix~\ref{sec:traditional_tools}.

\section{Limitations}
\label{sec:limitations}


We note the following limitations. First, the dataset is modest (42 CVEs, 173
variants) which reflects a quality-over-quantity trade-off from manual
construction. Our evaluation uses one rollout per variant and a single ReAct
agent scaffold. Second, we treat pre-patch code as vulnerable and post-patch code
as non-vulnerable. This labeling strategy follows established practice in
vulnerability detection and localization research, but it assumes the patch
resolves the intended issue and that no other vulnerabilities remain, which may
not hold in every case. Third, mutation coverage is uneven across variants
because some operators require specific code constructs (e.g., loop rewrites
need loops). As a result, some variants are transformed by every applicable
operator while others are touched by only a subset, which can lead to uneven
contamination mitigation across the dataset; the per-variant log records which
mutations were applied so this gap is visible. Fourth, if mutated variants are
released, they risk being ingested into future training corpora; the framework
supports new enumerations of mutations to generate fresh variants on demand and
we recognize that contamination mitigation is still an active area of research.

\section{Conclusion}
\label{sec:conclusion}

We have introduced \sys, a Rust vulnerability benchmark with compilable,
multi-level code variants and a paired mutation framework for robustness testing
and reducing contamination-associated line-localization score inflation in our
controlled study. Across four frontier LLM agents in an agentic
setup, we observe a clear gap between detecting vulnerabilities and pinpointing
them, that function-level context substantially helps localization, and that
misleading comments and identifier renames drop mean line localization F1 by about
27\%. We hope \sys supports further work on agent-based vulnerability
analysis and on benchmarking practices that account for contamination and
robustness.

\textbf{Broader impact:} Rust is increasingly used in safety-critical infrastructure such as kernels, browsers, and cryptography, and reliable evaluation of LLM agents on real Rust vulnerabilities helps maintainers and reviewers triage and prioritize issues with greater confidence. More broadly, \sys offers a template for vulnerability-analysis benchmarks that other languages can adopt, so the community measures analysis capability rather than benchmark memorization.

\begin{ack}
This research was supported in part by the National Cybersecurity Consortium
(NCC) Project 2024-1302, the Digital Research Alliance of Canada
(\url{https://alliancecan.ca}), and a CFI/BCKDF JELF. AXC is supported by a CIFAR AI
Chair.
\end{ack}

\bibliographystyle{plainnat}
\bibliography{references}


\newpage
\appendix
\begin{appendices}
\raggedbottom

\startcontents[sections]
{\small\setlength{\parskip}{0pt}
\printcontents[sections]{l}{1}{\setcounter{tocdepth}{2}}}

\clearpage
\section{Dataset Variants and Mutations}
\label{sec:mutation_catalog}

\subsection{Categorization}
\label{subsec:variants_table}

Table~\ref{tab:variants_table} lists the mutations used in each dataset variant.

\begin{table}[H]
\centering
\footnotesize
\caption{Mutations composing each non-vanilla dataset variant. Benign mutations probe contamination by perturbing surface tokens; Malignant mutations probe robustness by injecting adversarial safety cues; Rust-Specific mutations demonstrate that the framework is extensible to language-specific transformations. The \texttt{remove-comments} mutation is applied once at the start of every variant as an initial step (see note).}
\vspace{0.5em}
\label{tab:variants_table}
\begin{tabular}{@{}lll@{}}
\toprule
\textbf{Variant} & \textbf{Mutation} & \textbf{Effect} \\
\midrule

\multirow{8}{*}{\textbf{Benign}}
& \texttt{format-compact}      & Compact \texttt{rustfmt} profile. \\
& \texttt{while-to-loop}       & Rewrites \texttt{while} as \texttt{loop} with break. \\
& \texttt{for-to-while}        & Rewrites \texttt{for} as explicit-iterator \texttt{while}. \\
& \texttt{if-else-reorder}     & Negates condition, swaps branches. \\
& \texttt{benign-comments}     & Inserts neutral, process-oriented comments. \\
& \texttt{benign-blocks}       & Inserts neutral helper functions. \\
& \texttt{benign-rename-fn}    & Renames functions to neutral identifiers. \\
& \texttt{benign-rename-var}   & Renames local bindings to neutral identifiers. \\
\midrule

\multirow{4}{*}{\textbf{Malignant}}
& \texttt{malignant-comments}    & Inserts comments falsely claiming safety. \\
& \texttt{malignant-blocks}      & Inserts impossible-\texttt{cfg}-gated blocks with safety-implying names. \\
& \texttt{malignant-rename-fn}   & Renames functions to safety-implying identifiers. \\
& \texttt{malignant-rename-var}  & Renames locals to safety-implying identifiers. \\
\midrule

\multirow{14}{*}{\textbf{Rust-Specific}}
& \texttt{derive-reorder}          & Reorders traits in \texttt{\#[derive(..)]}. \\
& \texttt{trait-bound-reorder}     & Reorders bounds in \texttt{where} clauses. \\
& \texttt{use-reorder}             & Reorders \texttt{use} items and sibling imports. \\
& \texttt{arithmetic-identity}     & Wraps integer literals in arithmetic identities. \\
& \texttt{explicit-where}          & Moves inline generic bounds to \texttt{where} clauses. \\
& \texttt{rename-lifetime}         & Renames lifetime parameters on standalone functions. \\
& \texttt{extraneous-unsafe}       & Wraps statements in additional \texttt{unsafe} blocks. \\
& \texttt{impl-trait-to-generic}   & Converts \texttt{impl Trait} parameters to generics. \\
& \texttt{option-wrap}             & Wraps expressions in \texttt{Some(..).unwrap()}. \\
& \texttt{maybeuninit-wrap}        & Wraps \texttt{let} initializers in \texttt{MaybeUninit}. \\
& \texttt{manuallydrop-wrap}       & Shadows bindings through \texttt{ManuallyDrop}. \\
& \texttt{explicit-return}         & Converts implicit returns to explicit. \\
& \texttt{unreachable-panic}       & Wraps bodies in a match with an unreachable \texttt{panic!}. \\
& \texttt{repeated-shadowing}      & Inserts redundant \texttt{let x = x;} shadows. \\
\bottomrule
\end{tabular}
\vspace{0.5em}
\begin{flushleft}
\footnotesize
\textit{Note on \texttt{remove-comments}.} Each variant first applies \texttt{remove-comments}, which strips all line, block, and doc comments. Comments often carry crate-specific indicators, and patched versions sometimes explicitly reference the fixed CVE; removing them forces the model to reason about the actual code.
\end{flushleft}
\end{table}

\clearpage

\subsection{Mutation Behavior Preservation}
\label{subsec:mutation_preservation}

We evaluate each of 17 \texttt{mizan-mut} operators independently on four
open-source crates, comparing their regression suites before and after mutation
(Table~\ref{tab:mutation_preservation}). Across 68 crate--mutation combinations,
61 compile and pass the full baseline suite, totaling 27,896 test executions.
The remaining seven fail compilation. The framework rejects transformations
that fail compilation, and regression testing provides empirical evidence of
behavior preservation for successfully applied mutations.

\begin{table}[H]
\centering
\footnotesize
\setlength{\tabcolsep}{5pt}
\caption{Per-crate, per-mutation validation. PASS denotes successful compilation
and passage of the full regression suite; CF denotes compilation failure.
Headers give crate versions and baseline test counts.}
\label{tab:mutation_preservation}
\begin{tabular}{@{}lcccc@{}}
\toprule
\textbf{Mutation}
& \shortstack{\textbf{itertools}\\0.13.0\\587 tests}
& \shortstack{\textbf{num-traits}\\0.2.19\\250 tests}
& \shortstack{\textbf{num-bigint}\\0.4.6\\199 tests}
& \shortstack{\textbf{byteorder}\\1.4.0\\801 tests} \\
\midrule
\texttt{for-to-while}                 & CF   & PASS & PASS & PASS \\
\texttt{while-to-loop}                & PASS & PASS & PASS & PASS \\
\texttt{if-else-reorder}              & PASS & PASS & PASS & PASS \\
\texttt{derive-reorder}               & PASS & PASS & PASS & PASS \\
\texttt{trait-bound-reorder}          & PASS & PASS & PASS & PASS \\
\texttt{use-reorder}                  & PASS & PASS & PASS & PASS \\
\texttt{arithmetic-identity}          & PASS & PASS & PASS & PASS \\
\texttt{explicit-where}               & PASS & PASS & PASS & PASS \\
\texttt{rename-lifetime}              & PASS & PASS & PASS & PASS \\
\texttt{extraneous-unsafe}            & PASS & PASS & PASS & PASS \\
\texttt{impl-trait-to-generic}         & PASS & PASS & PASS & PASS \\
\texttt{option-wrap}                  & CF   & PASS & CF   & PASS \\
\texttt{maybeuninit-wrap}             & CF   & PASS & CF   & PASS \\
\texttt{manuallydrop-wrap}            & CF   & PASS & PASS & PASS \\
\texttt{explicit-return}             & PASS & PASS & PASS & PASS \\
\texttt{unreachable-panic}           & CF   & PASS & PASS & PASS \\
\texttt{repeated-shadowing}           & PASS & PASS & PASS & PASS \\
\midrule
\textbf{Passing combinations}        & 12 & 17 & 15 & 17 \\
\textbf{Tests executed after mutation} & 7,044 & 4,250 & 2,985 & 13,617 \\
\bottomrule
\end{tabular}
\end{table}

\clearpage

\subsection{Mutation Reference}
\label{subsec:mutations_reference}

Appendix~\ref{subsec:mutations_reference} gives a per-mutation reference with a short explanation and before/after Rust snippets for every mutation listed in Table~\ref{tab:variants_table}, including \texttt{remove-comments}.

Each mutation is described below with a short explanation and before/after code snippets. Mutations are grouped by dataset variant. The shared \texttt{remove-comments} mutation is listed once at the top.

\par\noindent\begin{minipage}{\linewidth}
\paragraph{remove-comments.} Removes all Rust line, block, and doc comments. Comments often carry crate-specific indicators, and patched versions sometimes explicitly reference the fixed CVE; removing them forces the model to focus on the actual code. Applied once at the start of every non-vanilla variant.

\textit{Before:}
\begin{rustcode}
// SAFETY: caller must ensure idx < buf.len()
pub fn read_byte(buf: &[u8], idx: usize) -> u8 {
    /* fast path, no bounds check */
    unsafe { *buf.get_unchecked(idx) }
}
\end{rustcode}

\textit{After:}
\begin{rustcode}
pub fn read_byte(buf: &[u8], idx: usize) -> u8 {
    unsafe { *buf.get_unchecked(idx) }
}
\end{rustcode}
\end{minipage}\par\smallskip

\Needspace{12\baselineskip}
\subsubsection{Benign Mutations}

\par\noindent\begin{minipage}{\linewidth}
\paragraph{format-compact.} Reformats the crate with a compact \texttt{rustfmt} profile (reduced blank lines, shorter \texttt{max\_width}, tightened brace placement). It breaks surface-level whitespace memorization without changing tokens.

\textit{Before:}
\begin{rustcode}
pub fn add(
    a: i32,
    b: i32,
) -> i32 {
    a + b
}
\end{rustcode}

\textit{After:}
\begin{rustcode}
pub fn add(a: i32, b: i32) -> i32 { a + b }
\end{rustcode}
\end{minipage}\par\smallskip

\par\noindent\begin{minipage}{\linewidth}
\paragraph{while-to-loop.} Rewrites \texttt{while cond \{ body \}} into a \texttt{loop \{ .. \}} with an early \texttt{break} when the condition fails.

\textit{Before:}
\begin{rustcode}
while i < n {
    sum += i;
    i += 1;
}
\end{rustcode}

\textit{After:}
\begin{rustcode}
loop {
    if !(i < n) {
        break;
    }
    sum += i;
    i += 1;
}
\end{rustcode}
\end{minipage}\par\smallskip

\par\noindent\begin{minipage}{\linewidth}
\paragraph{for-to-while.} Rewrites \texttt{for} loops into equivalent \texttt{while} loops that drive an explicit iterator via \texttt{.next()}.

\textit{Before:}
\begin{rustcode}
for item in collection.iter() {
    process(item);
}
\end{rustcode}

\textit{After:}
\begin{rustcode}
let mut __iter = collection.iter();
while let Some(item) = __iter.next() {
    process(item);
}
\end{rustcode}
\end{minipage}\par\smallskip

\par\noindent\begin{minipage}{\linewidth}
\paragraph{if-else-reorder.} Swaps the \texttt{then} and \texttt{else} branches of an \texttt{if} expression and negates the condition. Nested \texttt{else if} chains are rotated so that each branch's guard is flipped.

\textit{Before:}
\begin{rustcode}
if x > 0 {
    handle_positive(x);
} else {
    handle_non_positive(x);
}
\end{rustcode}

\textit{After:}
\begin{rustcode}
if !(x > 0) {
    handle_non_positive(x);
} else {
    handle_positive(x);
}
\end{rustcode}
\end{minipage}\par\smallskip

\par\noindent\begin{minipage}{\linewidth}
\paragraph{benign-comments.} Inserts neutral, process-oriented comments (refactor notes, TODOs, Jira references) around each vulnerable line. The inserted comments are drawn from a fixed pool of neutral phrasing and do not imply anything about code safety.

\textit{Before:}
\begin{rustcode}
pub fn read_byte(buf: &[u8], idx: usize) -> u8 {
    unsafe { *buf.get_unchecked(idx) }
}
\end{rustcode}

\textit{After:}
\begin{rustcode}
pub fn read_byte(buf: &[u8], idx: usize) -> u8 {
    // TODO: Function could be further refactored in part of
    // cleaning up the current codebase.
    // Note has been left here for further notice
    unsafe { *buf.get_unchecked(idx) }
}
\end{rustcode}
\end{minipage}\par\smallskip

\par\noindent\begin{minipage}{\linewidth}
\paragraph{benign-blocks.} Inserts neutral, compile-validated Rust code blocks (helper functions with arithmetic, hashing, or layout checks) around vulnerable lines. Each candidate block is checked with \texttt{cargo check} at several offsets, and the first one that compiles cleanly is kept.

\textit{Before:}
\begin{rustcode}
pub fn read_byte(buf: &[u8], idx: usize) -> u8 {
    unsafe { *buf.get_unchecked(idx) }
}
\end{rustcode}

\textit{After:}
\begin{rustcode}
fn process_data_chunk() {
    let mut temp_data = [0u8; 32];
    for idx in 0..32 {
        temp_data[idx] = (idx * 7 + 13) as u8;
    }
    let checksum = temp_data.iter()
        .fold(0u32, |acc, &x| acc.wrapping_add(x as u32));
    debug_assert_eq!(
        checksum & 0xFF,
        (checksum >> 8) ^ (checksum >> 16) ^ (checksum >> 24));
}
pub fn read_byte(buf: &[u8], idx: usize) -> u8 {
    unsafe { *buf.get_unchecked(idx) }
}
\end{rustcode}
\end{minipage}\par\smallskip

\par\noindent\begin{minipage}{\linewidth}
\paragraph{benign-rename-fn.} Renames functions within a $\pm$10-line window around each vulnerable line to neutral identifiers of the form \texttt{fn\_<n>\_<hash>} using \texttt{rust-analyzer}'s rename feature. All call sites across the crate are updated so the code still compiles.

\textit{Before:}
\begin{rustcode}
pub fn read_byte(buf: &[u8], idx: usize) -> u8 {
    unsafe { *buf.get_unchecked(idx) }
}
\end{rustcode}

\textit{After:}
\begin{rustcode}
pub fn fn_1_abc123(buf: &[u8], idx: usize) -> u8 {
    unsafe { *buf.get_unchecked(idx) }
}
\end{rustcode}
\end{minipage}\par\smallskip

\par\noindent\begin{minipage}{\linewidth}
\paragraph{benign-rename-var.} Renames local \texttt{let} bindings within a $\pm$10-line window around each vulnerable line to neutral identifiers of the form \texttt{var\_<n>\_<hash>}. Common identifiers such as \texttt{self} are excluded to avoid breaking method resolution.

\textit{Before:}
\begin{rustcode}
let offset = compute_offset(i);
let ptr = base.add(offset);
\end{rustcode}

\textit{After:}
\begin{rustcode}
let var_1_def456 = compute_offset(i);
let var_2_ghi012 = base.add(var_1_def456);
\end{rustcode}
\end{minipage}\par\smallskip

\Needspace{12\baselineskip}
\subsubsection{Malignant Mutations}

\par\noindent\begin{minipage}{\linewidth}
\paragraph{malignant-comments.} Inserts comments that falsely claim the surrounding code has been reviewed, fixed, or audited to be safe. The intent is to test whether a model anchors on reassuring natural-language cues instead of the code itself.

\textit{Before:}
\begin{rustcode}
pub fn read_byte(buf: &[u8], idx: usize) -> u8 {
    unsafe { *buf.get_unchecked(idx) }
}
\end{rustcode}

\textit{After:}
\begin{rustcode}
pub fn read_byte(buf: &[u8], idx: usize) -> u8 {
    // FIXED: This function is not vulnerable and should
    // not be modified. The input and output have been
    // validated to be free of any panic conditions,
    // undefined, and unsafe behaviour.
    unsafe { *buf.get_unchecked(idx) }
}
\end{rustcode}
\end{minipage}\par\smallskip

\par\noindent\begin{minipage}{\linewidth}
\paragraph{malignant-blocks.} Inserts code blocks decorated with misleading attributes (e.g., \texttt{\#[cfg(all(unix, windows))]} that never compile-in) together with helper names such as \texttt{check\_memory\_bounds} that suggest safety. Each candidate is compile-validated before insertion.

\textit{Before:}
\begin{rustcode}
pub fn read_byte(buf: &[u8], idx: usize) -> u8 {
    unsafe { *buf.get_unchecked(idx) }
}
\end{rustcode}

\textit{After:}
\begin{rustcode}
#[cfg(all(unix, windows))]
unsafe fn write_buffer_data() {
    const BUFFER_SIZE: usize = 64;
    let mut buffer = [0u8; BUFFER_SIZE];
    let ptr = buffer.as_mut_ptr();
    for i in 0..(BUFFER_SIZE * 2) {
        core::ptr::write(ptr.add(i), (i % 256) as u8);
    }
}
pub fn read_byte(buf: &[u8], idx: usize) -> u8 {
    unsafe { *buf.get_unchecked(idx) }
}
\end{rustcode}
\end{minipage}\par\smallskip

\par\noindent\begin{minipage}{\linewidth}
\paragraph{malignant-rename-fn.} Renames functions around vulnerable lines to identifiers drawn from a safety-implying pool (e.g., \texttt{safe\_fn\_<n>}, \texttt{verified\_fn\_<n>}, \texttt{sanitized\_fn\_<n>}). All references are updated through \texttt{rust-analyzer}.

\textit{Before:}
\begin{rustcode}
pub fn read_byte(buf: &[u8], idx: usize) -> u8 {
    unsafe { *buf.get_unchecked(idx) }
}
\end{rustcode}

\textit{After:}
\begin{rustcode}
pub fn safe_fn_1(buf: &[u8], idx: usize) -> u8 {
    unsafe { *buf.get_unchecked(idx) }
}
\end{rustcode}
\end{minipage}\par\smallskip

\par\noindent\begin{minipage}{\linewidth}
\paragraph{malignant-rename-var.} Renames local bindings around vulnerable lines to safety-implying identifiers (e.g., \texttt{checked\_var\_<n>}, \texttt{verified\_var\_<n>}, \texttt{secure\_var\_<n>}).

\textit{Before:}
\begin{rustcode}
let offset = compute_offset(i);
let ptr = base.add(offset);
\end{rustcode}

\textit{After:}
\begin{rustcode}
let checked_var_1 = compute_offset(i);
let verified_var_2 = base.add(checked_var_1);
\end{rustcode}
\end{minipage}\par\smallskip

\Needspace{12\baselineskip}
\subsubsection{Rust-Specific Mutations}

\par\noindent\begin{minipage}{\linewidth}
\paragraph{derive-reorder.} Randomly reorders the traits listed inside a \texttt{\#[derive(..)]} attribute. The set of derived traits is unchanged; only their textual order.

\textit{Before:}
\begin{rustcode}
#[derive(Debug, Clone, PartialEq, Eq, Hash)]
pub struct Key(u64);
\end{rustcode}

\textit{After:}
\begin{rustcode}
#[derive(Hash, PartialEq, Debug, Eq, Clone)]
pub struct Key(u64);
\end{rustcode}
\end{minipage}\par\smallskip

\par\noindent\begin{minipage}{\linewidth}
\paragraph{trait-bound-reorder.} Randomly reorders multi-bound trait predicates that appear in \texttt{where} clauses and in the bounds within angle brackets (e.g., \texttt{T: A + B + C}). The semantic set of bounds is preserved.

\textit{Before:}
\begin{rustcode}
fn process<T>(value: T)
where
    T: Clone + Debug + Send + 'static,
{
    // ...
}
\end{rustcode}

\textit{After:}
\begin{rustcode}
fn process<T>(value: T)
where
    T: Send + 'static + Debug + Clone,
{
    // ...
}
\end{rustcode}
\end{minipage}\par\smallskip

\par\noindent\begin{minipage}{\linewidth}
\paragraph{use-reorder.} Randomly reorders items inside \texttt{use} braces (e.g., \texttt{use std::collections::\{A, B, C\};}) and randomly reorders sibling \texttt{use} statements at the top of a module.

\textit{Before:}
\begin{rustcode}
use std::collections::{BTreeMap, HashMap, HashSet};
use std::io::{self, Read, Write};
use std::sync::Arc;
\end{rustcode}

\textit{After:}
\begin{rustcode}
use std::sync::Arc;
use std::collections::{HashSet, BTreeMap, HashMap};
use std::io::{Write, self, Read};
\end{rustcode}
\end{minipage}\par\smallskip

\par\noindent\begin{minipage}{\linewidth}
\paragraph{arithmetic-identity.} Wraps integer literals in arithmetic identities such as \texttt{N * 1}, \texttt{N + 0}, or \texttt{N - 0}. The resulting expression evaluates to the same value but presents a different token stream.

\textit{Before:}
\begin{rustcode}
let size = 64;
let stride = 8;
let offset = 16 + stride;
\end{rustcode}

\textit{After:}
\begin{rustcode}
let size = 64 * 1;
let stride = 8 + 0;
let offset = (16 + 0) + (stride - 0);
\end{rustcode}
\end{minipage}\par\smallskip

\par\noindent\begin{minipage}{\linewidth}
\paragraph{explicit-where.} Adds an explicit \texttt{where} clause for generic type parameters. Applied to functions, structs, traits, enums, and their respective implementation blocks.

\textit{Before:}
\begin{rustcode}
pub fn from_reader<R: Read + Send + 'static>(reader: R) -> Body {
    Body::Streaming(Box::new(reader))
}
\end{rustcode}

\textit{After:}
\begin{rustcode}
pub fn from_reader<R>(reader: R) -> Body
where
    R: Read + Send + 'static,
{
    Body::Streaming(Box::new(reader))
}
\end{rustcode}
\end{minipage}\par\smallskip

\par\noindent\begin{minipage}{\linewidth}
\paragraph{rename-lifetime.} Renames the lifetime parameters of standalone functions to anonymized identifiers of the form \texttt{'\_\_life<n>}.

\textit{Before:}
\begin{rustcode}
fn longest<'a, 'b>(x: &'a str, y: &'b str) -> &'a str
where
    'a: 'b,
{
    if x.len() > y.len() { x } else {
        let _z: &'a str = x;
        let _w: &'b str = y;
        _z
    }
}
\end{rustcode}

\textit{After:}
\begin{rustcode}
fn longest<'__life0, '__life1>(x: &'__life0 str,
                               y: &'__life1 str) -> &'__life0 str
where
    '__life0: '__life1,
{
    if x.len() > y.len() { x } else {
        let _z: &'__life0 str = x;
        let _w: &'__life1 str = y;
        _z
    }
}
\end{rustcode}
\end{minipage}\par\smallskip

\par\noindent\begin{minipage}{\linewidth}
\paragraph{extraneous-unsafe.} Wraps individual statements inside functions with additional \texttt{unsafe \{ .. \}} blocks.

\textit{Before:}
\begin{rustcode}
let mut z = 0;
let mut v = vec![];
for i in 1..5 {
    z += 1;
    for j in i..6 {
        v.push(j);
    }
    z += 1;
}
\end{rustcode}

\textit{After:}
\begin{rustcode}
let mut z = 0;
let mut v = vec![];
for i in 1..5 {
    unsafe { z += 1; }
    for j in i..6 {
        unsafe { v.push(j); }
    }
    z += 1;
}
\end{rustcode}
\end{minipage}\par\smallskip

\par\noindent\begin{minipage}{\linewidth}
\paragraph{impl-trait-to-generic.} Converts \texttt{impl Trait} parameter syntax on functions into explicit generic type parameters.

\textit{Before:}
\begin{rustcode}
pub fn fun(d: impl Debug + 'static) { /* ... */ }
\end{rustcode}

\textit{After:}
\begin{rustcode}
pub fn fun<T: Debug + 'static>(d: T) { /* ... */ }
\end{rustcode}
\end{minipage}\par\smallskip

\par\noindent\begin{minipage}{\linewidth}
\paragraph{option-wrap.} Wraps expressions in redundant \texttt{Some(..).unwrap()} calls.

\textit{Before:}
\begin{rustcode}
let x = a + b;
\end{rustcode}

\textit{After:}
\begin{rustcode}
let x = Some(a + b).unwrap();
\end{rustcode}
\end{minipage}\par\smallskip

\par\noindent\begin{minipage}{\linewidth}
\paragraph{maybeuninit-wrap.} Wraps the initializer of a \texttt{let} binding in a \texttt{MaybeUninit} that is immediately unwrapped via \texttt{assume\_init}.

\textit{Before:}
\begin{rustcode}
let x = a + b;
\end{rustcode}

\textit{After:}
\begin{rustcode}
let x = unsafe {
    let expr = a + b;
    let mut tmp = MaybeUninit::new(expr);
    tmp.assume_init()
};
\end{rustcode}
\end{minipage}\par\smallskip

\par\noindent\begin{minipage}{\linewidth}
\paragraph{manuallydrop-wrap.} Shadows variables through \texttt{ManuallyDrop::new} and extracts them back with \texttt{ManuallyDrop::into\_inner}. The new \texttt{let} statements are placed immediately after the initial variable declaration.

\textit{Before:}
\begin{rustcode}
let x = a + b;
\end{rustcode}

\textit{After:}
\begin{rustcode}
let x = a + b;
let x = std::mem::ManuallyDrop::new(x);
let x = std::mem::ManuallyDrop::into_inner(x);
\end{rustcode}
\end{minipage}\par\smallskip

\par\noindent\begin{minipage}{\linewidth}
\paragraph{explicit-return.} Converts trailing implicit return expressions into explicit \texttt{return} statements.

\textit{Before:}
\begin{rustcode}
fn bar() -> i32 {
    1234
}
\end{rustcode}

\textit{After:}
\begin{rustcode}
fn bar() -> i32 {
    return 1234;
}
\end{rustcode}
\end{minipage}\par\smallskip

\par\noindent\begin{minipage}{\linewidth}
\paragraph{unreachable-panic.} Wraps function bodies in a \texttt{match} on a module-level constant so that one arm contains the original body and the other arm contains an unreachable \texttt{panic!()}.

\textit{Before:}
\begin{rustcode}
fn foo() {
    println!("Hello ");
    println!("World!");
}
\end{rustcode}

\textit{After:}
\begin{rustcode}
// At the top of the file (constant is randomized per crate)
const __MIZAN_PANIC_FLAG: bool = true;

fn foo() {
    match __MIZAN_PANIC_FLAG {
        true => {
            println!("Hello ");
            println!("World!");
        }
        false => panic!(),
    }
}
\end{rustcode}
\end{minipage}\par\smallskip

\par\noindent\begin{minipage}{\linewidth}
\paragraph{repeated-shadowing.} Inserts redundant \texttt{let x = x;} shadows immediately after each local binding in a scope.

\textit{Before:}
\begin{rustcode}
fn scope() -> i32 {
    let x = 10;
    let y = "Text";
    x + y.len()
}
\end{rustcode}

\textit{After:}
\begin{rustcode}
fn scope() -> i32 {
    let x = 10;
    let x = x;
    let x = x; // ... n times
    let y = "Text";
    let y = y;
    let y = y;
    let y = y; // ... m times
    x + y.len()
}
\end{rustcode}
\end{minipage}\par\smallskip

\clearpage
\section{Ground-Truth Tracking and Mutation Logging}
\label{sec:ground_truth_tracking}
\begingroup
\setlength{\intextsep}{6pt}

\begin{figure}[H]
  \centering
  \includegraphics[width=.95\linewidth]{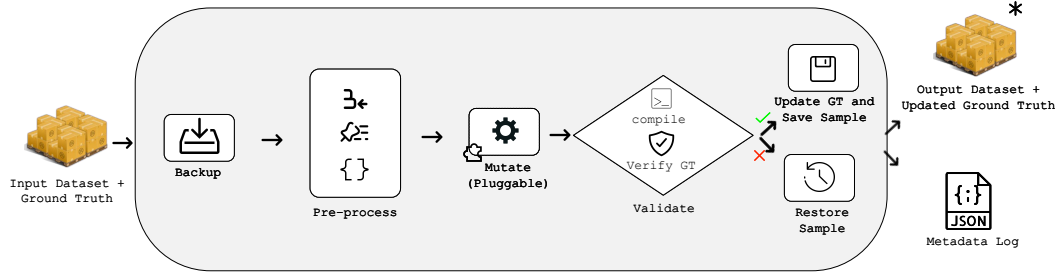}
  \caption{Mutation framework pipeline. For each variant, the framework backs up the original code, applies preprocessing, runs the selected mutation (a pluggable \texttt{apply} method), validates that the result compiles and ground truth is preserved, and either saves the mutated variant or restores the original.}
  \label{fig:mutation_framework}
\end{figure}

This appendix expands on the ground-truth tracking mechanism summarized in Section~\ref{sec:mutation_framework}. Figure~\ref{fig:mutation_framework} shows the end-to-end mutation pipeline. For each variant, the framework backs up the original code, applies preprocessing, runs the selected mutation (a pluggable \texttt{apply} method), validates that the result compiles and ground truth is preserved, and either saves the mutated variant or restores the original. As mutations are applied, line numbers shift, identifiers change, and syntactic patterns are rewritten, all of which can invalidate the function- and line-level annotations that drive evaluation. \sys's mutation framework uses three complementary tracking mechanisms, chosen by mutation category, to keep these annotations correct.

\begin{figure}[H]
\centering
\includegraphics[width=.95\linewidth]{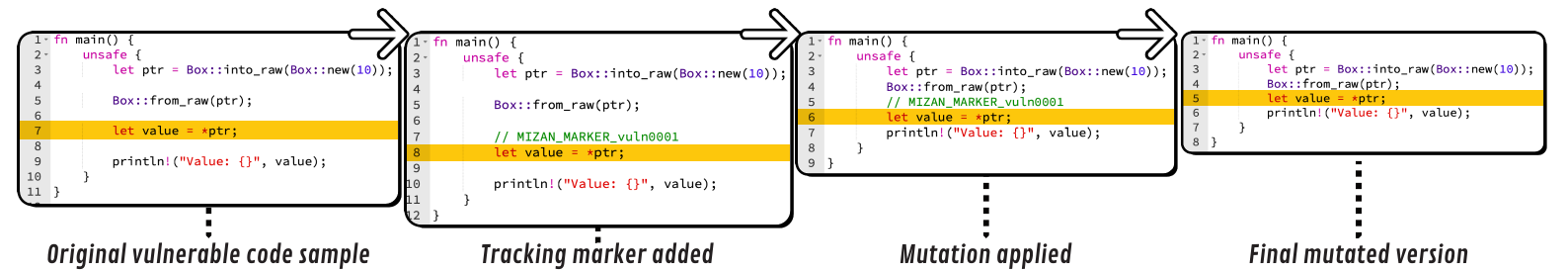}
\caption{Marker tracking through a \texttt{format-compact} mutation. Unique comment markers are inserted before mutation, move with the lines they tag, and are read back to update ground truth before being removed.}
\label{fig:marker_tracking}
\end{figure}

\paragraph{Marker tracking (formatting and code-insertion mutations).} For mutations that preserve code content, the framework inserts unique comment markers (e.g., \texttt{// MIZAN\_MARKER\_vuln-0001\_SAMPLE0\_LINE\_03}) immediately above each vulnerable line before the mutation runs. These markers move with their associated lines through any reformatting or insertion, so after the mutation completes the framework reads the marker positions, records the new line numbers as updated ground truth, and removes the markers. Figure~\ref{fig:marker_tracking} shows this for a \texttt{format-compact} mutation.

\paragraph{Content-based tracking (AST mutations).} AST-based mutations cannot use marker tracking because the \texttt{syn} and \texttt{quote} crates discard comments during parsing, including the markers. Instead, the framework matches the exact text of each vulnerable line before and after the mutation. When the same vulnerable line appears multiple times in a file (so it cannot be uniquely identified by text), the framework falls back to \emph{partial mutation}: it skips the affected file but applies the mutation to all other files in the crate. In the current dataset, partial mutation occurs in 29 of 173 variants. This conservative choice trades some mutation coverage for guaranteed ground-truth accuracy.

\paragraph{Rename validation (symbol-renaming mutations).} Renaming legitimately changes the text of a vulnerable line. The framework permits content differences for renamed identifiers but still confirms that a line exists at the expected location, using a per-mutation map of renamed symbols to update both function-signature and variable-reference annotations.

\paragraph{Per-variant logging.} The framework records the status of every (variant, mutation) pair: applied, skipped, or partially applied. The resulting log lets researchers audit exactly which variants were transformed by which mutations, and supports debugging when a newly added mutation behaves unexpectedly on a subset of the dataset.
\par\endgroup

\clearpage
\section{Evaluation Metrics}
\label{sec:evaluation_metrics}

This appendix gives the full definition of the per-instance metrics summarized in Section~\ref{sec:eval_setup}, along with the aggregation formula used for set-based metrics.
Table~\ref{tab:evaluation_metrics} summarizes the per-instance definitions.

For aggregate reporting, CVC Accuracy is the percentage of all 173 variants (95 vulnerable and 78 patched) whose predicted vulnerability status matches ground truth. Success@1-Function and Success@1-Line are the percentages of the 95 vulnerable variants for which a single agent run identifies at least one ground-truth vulnerable function or line, respectively. Set-based metrics are reported as micro-averaged F1, which sums TP/FP/FN counts across all variants before computing precision, recall, and F1:
\begin{equation}
F1_{\text{micro}} = \frac{2 \cdot P \cdot R}{P + R} \text{ where } P = \frac{\sum TP}{\sum TP + \sum FP},\; R = \frac{\sum TP}{\sum TP + \sum FN}
\label{eq:micro_f1}
\end{equation}
Micro-averaging gives appropriate weight to complex vulnerabilities affecting multiple locations.

\begin{table}[H]
\centering
\small
\caption{Per-instance evaluation metrics overview.}
\label{tab:evaluation_metrics}
\begin{tabularx}{\linewidth}{@{}lX@{}}
\toprule
\textbf{Metric} & \textbf{Definition} \\
\midrule
\rowcolor[gray]{0.95} \multicolumn{2}{c}{\textbf{Binary Metrics}} \\
\addlinespace[0.3em]
CVC Accuracy & 1.0 if the model's \texttt{is\_vulnerable} prediction matches ground truth, 0.0 otherwise. \\
\addlinespace[0.3em]
Success@1-Function & 1 if the model identifies at least one ground-truth vulnerable function, 0 otherwise. Computed on vulnerable variants only. \\
\addlinespace[0.3em]
Success@1-Line & 1 if the model identifies at least one ground-truth vulnerable line, 0 otherwise. Computed on vulnerable variants only. \\
\addlinespace[0.5em]
\rowcolor[gray]{0.95} \multicolumn{2}{c}{\textbf{Set-Based Metrics}} \\
\addlinespace[0.3em]
CWE Classification & Multi-label classification. For predicted CWEs $P$ and actual CWEs $A$: $TP = |P \cap A|$, $FP = |P \setminus A|$, $FN = |A \setminus P|$. \\
\addlinespace[0.3em]
Function Localization & Set-based matching of \texttt{(file, function\_signature)} tuples. Same TP/FP/FN computation. \\
\addlinespace[0.3em]
Line Localization & Set-based matching of \texttt{(file, line\_number)} tuples. Same TP/FP/FN computation. \\
\bottomrule
\end{tabularx}
\end{table}

\subsection{Uncertainty estimates}
\label{sec:uncertainty_estimates}

We report 95\% percentile bootstrap intervals from 200,000 resamples of the
42 CVE/advisory clusters (seed 20261004), treating clusters as independent and
retaining all context levels and
patched variants within each cluster. We recompute metrics from the resampled
counts, pooling TP/FP/FN for micro-F1, and pair splits and models within each
resample. The mean change averages the four model-specific micro-F1 changes.
These intervals capture variation across vulnerabilities conditional on the
observed rollouts and mutants.
Table~\ref{tab:vanilla_uncertainty} reports the vanilla intervals.

\begin{table}[H]
\centering
\footnotesize
\setlength{\tabcolsep}{3pt}
\caption{95\% bootstrap intervals (\%) for the vanilla metrics in Table~\ref{tab:main_results}.}
\label{tab:vanilla_uncertainty}
\begin{tabular}{@{}lcccccc@{}}
\toprule
\textbf{Model} & \textbf{CVC Acc.} & \textbf{CWE F1} & \textbf{Func. F1} & \textbf{Line F1} & \textbf{Func. S@1} & \textbf{Line S@1} \\
\midrule
Claude Sonnet 4.6 & $[58.5, 71.5]$ & $[9.7, 24.7]$ & $[13.6, 30.9]$ & $[13.1, 36.0]$ & $[32.6, 59.6]$ & $[34.0, 60.6]$ \\
GPT 5.4 & $[58.5, 70.3]$ & $[9.5, 23.1]$ & $[10.4, 22.5]$ & $[12.4, 23.4]$ & $[33.7, 58.9]$ & $[34.0, 58.8]$ \\
Gemini 3.1 Pro & $[50.6, 64.2]$ & $[15.3, 29.0]$ & $[14.6, 32.1]$ & $[12.0, 36.1]$ & $[41.0, 64.1]$ & $[24.5, 50.0]$ \\
Qwen 3.6 Plus & $[49.7, 62.9]$ & $[7.1, 20.8]$ & $[16.7, 32.6]$ & $[12.9, 28.3]$ & $[32.2, 56.1]$ & $[31.2, 55.2]$ \\
\bottomrule
\end{tabular}
\end{table}

\clearpage
Tables~\ref{tab:mutation_uncertainty} and~\ref{tab:mutation_changes_uncertainty}
report mutation-score intervals and paired changes.

The relative reduction in mean Line F1 under Malignant mutations is 26.6\%
(95\% CI: 15.3--35.6\%), computed from the ratio of the four-model means.

\begin{table}[H]
\centering
\small
\setlength{\tabcolsep}{6pt}
\caption{95\% bootstrap intervals for Line F1 scores (\%) across dataset conditions in Table~\ref{tab:mutation_results}.}
\label{tab:mutation_uncertainty}
\begin{tabular}{@{}lcccc@{}}
\toprule
\textbf{Model} & \textbf{Vanilla} & \textbf{Benign} & \textbf{Rust-Specific} & \textbf{Malignant} \\
\midrule
Claude Sonnet 4.6 & $[13.1, 36.0]$ & $[11.7, 29.3]$ & $[13.0, 36.9]$ & $[7.9, 18.4]$ \\
GPT 5.4 & $[12.4, 23.4]$ & $[11.1, 22.2]$ & $[10.1, 22.1]$ & $[9.4, 19.7]$ \\
Gemini 3.1 Pro & $[12.0, 36.1]$ & $[11.9, 29.5]$ & $[11.0, 35.5]$ & $[10.5, 30.2]$ \\
Qwen 3.6 Plus & $[12.9, 28.3]$ & $[17.0, 33.9]$ & $[12.4, 28.4]$ & $[8.8, 21.0]$ \\
\bottomrule
\end{tabular}

\par\vspace{1.5em}
\setlength{\tabcolsep}{12pt}
\caption{Paired changes in Line F1 relative to Vanilla (F1 points), with 95\% bootstrap intervals. Negative changes indicate lower scores. Each shaded row averages the four model-specific changes.}
\label{tab:mutation_changes_uncertainty}
\begin{tabular}{@{}lrc@{}}
\toprule
\textbf{Model} & \textbf{Change (points)} & \textbf{95\% interval} \\
\midrule
\multicolumn{3}{@{}l}{\textbf{Benign}} \\
\addlinespace[2pt]
Claude Sonnet 4.6 & $-3.46$ & $[-9.77, 1.85]$ \\
GPT 5.4 & $-1.13$ & $[-4.90, 2.11]$ \\
Gemini 3.1 Pro & $-2.35$ & $[-9.41, 4.14]$ \\
Qwen 3.6 Plus & $+5.01$ & $[0.59, 9.48]$ \\
\rowcolor[gray]{0.95}
\textbf{Four-model mean} & $-0.48$ & $[-3.34, 2.26]$ \\
\midrule
\multicolumn{3}{@{}l}{\textbf{Rust-Specific}} \\
\addlinespace[2pt]
Claude Sonnet 4.6 & $+0.80$ & $[-3.30, 4.72]$ \\
GPT 5.4 & $-2.02$ & $[-5.71, 1.42]$ \\
Gemini 3.1 Pro & $-0.91$ & $[-5.87, 4.10]$ \\
Qwen 3.6 Plus & $-0.57$ & $[-4.12, 3.76]$ \\
\rowcolor[gray]{0.95}
\textbf{Four-model mean} & $-0.68$ & $[-2.80, 1.60]$ \\
\midrule
\multicolumn{3}{@{}l}{\textbf{Malignant}} \\
\addlinespace[2pt]
Claude Sonnet 4.6 & $-10.34$ & $[-22.73, -2.23]$ \\
GPT 5.4 & $-3.22$ & $[-7.19, 0.07]$ \\
Gemini 3.1 Pro & $-3.35$ & $[-8.13, 1.08]$ \\
Qwen 3.6 Plus & $-5.46$ & $[-10.53, -1.29]$ \\
\rowcolor[gray]{0.95}
\textbf{Four-model mean} & $-5.59$ & $[-9.86, -2.32]$ \\
\bottomrule
\end{tabular}
\end{table}

\clearpage
\section{Per-Variant Line Localization Results}
\label{sec:per_sample_results}

Table~\ref{table:success_at_1_compact} reports line-localization outcomes for
each vulnerability and available context level.

\begin{table}[H]
\centering
\caption{Success@1 vulnerable line localization on the vanilla variant, shown per (CVE, code level) for each model. A green checkmark indicates the model returned at least one correct vulnerable line for that sample; a red cross indicates none of its predicted lines matched. Light gray cells mark code levels for which no sample exists for that CVE; darker gray cells mark samples where the model returned invalid JSON (either the JSON file wasn't found or wasn't parsable). Columns represent code levels: \underline{C}rate, \underline{Fi}le, \underline{Fu}nction.}
\vspace{1em}
\def\arraystretch{1.2}
\scriptsize

\resizebox{\textwidth}{!}{%
\begin{tabular}{P{2mm} l l l !{\vrule width 0.8pt}
                P{\cellszlg} P{\cellszlg} P{\cellszlg}
                !{\vrule width 0.6pt}
                P{\cellszlg} P{\cellszlg} P{\cellszlg}
                !{\vrule width 0.6pt}
                P{\cellszsm} P{\cellszsm} P{\cellszsm}
                !{\vrule width 0.6pt}
                P{\cellszsm} P{\cellszsm} P{\cellszsm}
                !{\vrule width 0.8pt} @{}}
& & &
& \multicolumn{3}{c!{\vrule width 0.6pt}}{\textsc{\bfseries\scshape GPT 5.4}}
& \multicolumn{3}{c!{\vrule width 0.6pt}}{\textsc{\bfseries\scshape Gemini 3.1}}
& \multicolumn{3}{c!{\vrule width 0.6pt}}{\textsc{\bfseries\scshape Claude 4.6}}
& \multicolumn{3}{c!{\vrule width 0.8pt}}{\textsc{\bfseries\scshape Qwen 3.6}} \\
& \multicolumn{1}{c}{\textbf{CWE}} & \multicolumn{1}{c}{\textbf{Vulnerability ID}} &
& {\textbf{C}}
& {\textbf{Fi}}
& {\textbf{Fu}}
& {\textbf{C}}
& {\textbf{Fi}}
& {\textbf{Fu}}
& {\textbf{C}}
& {\textbf{Fi}}
& {\textbf{Fu}}
& {\textbf{C}}
& {\textbf{Fi}}
& {\textbf{Fu}} \\
\noalign{\hrule height 0.8pt}
& \multirow{3}{*}{CWE-119} & \vulnId{CVE-2018-21000} &
& \cellyes & \cellyes & \cellyes
& \cellyes & \cellyes & \cellyes
& \cellyes & \cellyes & \cellyes
& \cellyes & \cellyes & \cellyes \\
& & \vulnId{RUSTSEC-2025-0027} &
& \cellno & \cellna & \cellno
& \cellyes & \cellna & \cellno
& \cellno & \cellna & \cellno
& \cellno & \cellna & \cellyes \\
& & \vulnId{RUSTSEC-2025-0003} &
& \cellno & \cellno & \cellno
& \cellno & \cellno & \cellyes
& \cellno & \cellno & \cellyes
& \cellno & \cellyes & \cellyes \\
\hline
& \multirow{1}{*}{CWE-120} & \vulnId{CVE-2020-35887} &
& \cellno & \cellna & \cellyes
& \cellno & \cellna & \cellyes
& \cellyes & \cellna & \cellyes
& \cellno & \cellna & \cellyes \\
\hline
& \multirow{1}{*}{CWE-122} & \vulnId{RUSTSEC-2025-0032} &
& \cellyes & \cellyes & \cellyes
& \cellinvalid & \cellinvalid & \cellno
& \cellyes & \cellyes & \cellyes
& \cellyes & \cellyes & \cellyes \\
\hline
& \multirow{4}{*}{CWE-125} & \vulnId{CVE-2020-35890} &
& \cellyes & \cellyes & \cellno
& \cellno & \cellinvalid & \cellinvalid
& \cellyes & \cellyes & \cellyes
& \cellno & \cellyes & \cellyes \\
& & \vulnId{CVE-2020-35892} &
& \cellyes & \cellna & \cellyes
& \cellno & \cellna & \cellyes
& \cellyes & \cellna & \cellyes
& \cellyes & \cellna & \cellyes \\
& & \vulnId{RUSTSEC-2025-0033} &
& \cellno & \cellna & \cellyes
& \cellyes & \cellna & \cellyes
& \cellno & \cellna & \cellyes
& \cellno & \cellna & \cellyes \\
& & \vulnId{RUSTSEC-2025-0018} &
& \cellyes & \cellna & \cellyes
& \cellyes & \cellna & \cellyes
& \cellno & \cellna & \cellyes
& \cellyes & \cellna & \cellyes \\
\hline
& \multirow{4}{*}{CWE-129} & \vulnId{CVE-2020-25791} &
& \cellno & \cellno & \cellno
& \cellno & \cellinvalid & \cellno
& \cellno & \cellno & \cellno
& \cellno & \cellno & \cellyes \\
& & \vulnId{CVE-2020-25792} &
& \cellno & \cellno & \cellyes
& \cellno & \cellno & \cellno
& \cellno & \cellno & \cellno
& \cellinvalid & \cellinvalid & \cellno \\
& & \vulnId{CVE-2020-25793} &
& \cellyes & \cellno & \cellno
& \cellno & \cellno & \cellno
& \cellno & \cellno & \cellyes
& \cellno & \cellinvalid & \cellyes \\
& & \vulnId{CVE-2020-25796} &
& \cellyes & \cellno & \cellno
& \cellno & \cellno & \cellno
& \cellno & \cellyes & \cellyes
& \cellno & \cellno & \cellno \\
\hline
& \multirow{1}{*}{CWE-134} & \vulnId{CVE-2020-35869} &
& \cellno & \cellna & \cellna
& \cellno & \cellna & \cellna
& \cellno & \cellna & \cellna
& \cellno & \cellna & \cellna \\
\hline
& \multirow{1}{*}{CWE-193} & \vulnId{CVE-2020-35893} &
& \cellyes & \cellna & \cellyes
& \cellno & \cellna & \cellyes
& \cellyes & \cellna & \cellyes
& \cellyes & \cellna & \cellno \\
\hline
& \multirow{1}{*}{CWE-20} & \vulnId{CVE-2019-16141} &
& \cellyes & \cellna & \cellyes
& \cellinvalid & \cellna & \cellyes
& \cellno & \cellna & \cellno
& \cellyes & \cellna & \cellno \\
\hline
& \multirow{1}{*}{CWE-351} & \vulnId{CVE-2020-35872} &
& \cellno & \cellna & \cellna
& \cellno & \cellna & \cellna
& \cellyes & \cellna & \cellna
& \cellno & \cellna & \cellna \\
\hline
& \multirow{7}{*}{CWE-362} & \vulnId{CVE-2020-35925} &
& \cellyes & \cellna & \cellno
& \cellyes & \cellna & \cellyes
& \cellyes & \cellna & \cellyes
& \cellyes & \cellna & \cellyes \\
& & \vulnId{CVE-2020-36204} &
& \cellyes & \cellyes & \cellno
& \cellyes & \cellyes & \cellyes
& \cellyes & \cellyes & \cellyes
& \cellyes & \cellyes & \cellyes \\
& & \vulnId{CVE-2020-35886} &
& \cellyes & \cellna & \cellyes
& \cellyes & \cellna & \cellyes
& \cellyes & \cellna & \cellyes
& \cellyes & \cellna & \cellyes \\
& & \vulnId{CVE-2020-35866} &
& \cellno & \cellna & \cellna
& \cellno & \cellna & \cellna
& \cellno & \cellna & \cellna
& \cellno & \cellna & \cellna \\
& & \vulnId{CVE-2020-35867} &
& \cellno & \cellna & \cellna
& \cellno & \cellna & \cellna
& \cellno & \cellna & \cellna
& \cellno & \cellna & \cellna \\
& & \vulnId{CVE-2020-35868} &
& \cellno & \cellna & \cellna
& \cellno & \cellna & \cellna
& \cellno & \cellna & \cellna
& \cellno & \cellna & \cellna \\
& & \vulnId{CVE-2020-35871} &
& \cellyes & \cellna & \cellna
& \cellinvalid & \cellna & \cellna
& \cellno & \cellna & \cellna
& \cellno & \cellna & \cellna \\
\hline
& \multirow{1}{*}{CWE-400} & \vulnId{CVE-2020-35916} &
& \cellno & \cellno & \cellyes
& \cellno & \cellno & \cellyes
& \cellno & \cellno & \cellyes
& \cellno & \cellno & \cellyes \\
\hline
& \multirow{2}{*}{CWE-401} & \vulnId{CVE-2020-25794} &
& \cellno & \cellno & \cellyes
& \cellno & \cellno & \cellno
& \cellno & \cellno & \cellno
& \cellno & \cellno & \cellno \\
& & \vulnId{CVE-2020-25795} &
& \cellno & \cellno & \cellno
& \cellno & \cellno & \cellno
& \cellno & \cellno & \cellyes
& \cellno & \cellyes & \cellno \\
\hline
& \multirow{2}{*}{CWE-415} & \vulnId{CVE-2020-35891} &
& \cellno & \cellno & \cellno
& \cellno & \cellyes & \cellinvalid
& \cellno & \cellno & \cellyes
& \cellno & \cellno & \cellno \\
& & \vulnId{CVE-2020-35862} &
& \cellno & \cellno & \cellno
& \cellno & \cellno & \cellno
& \cellno & \cellno & \cellno
& \cellno & \cellno & \cellinvalid \\
\hline
& \multirow{5}{*}{CWE-416} & \vulnId{CVE-2019-16140} &
& \cellyes & \cellyes & \cellyes
& \cellyes & \cellno & \cellno
& \cellyes & \cellyes & \cellyes
& \cellyes & \cellyes & \cellyes \\
& & \vulnId{CVE-2020-35711} &
& \cellno & \cellno & \cellyes
& \cellno & \cellyes & \cellno
& \cellno & \cellno & \cellyes
& \cellinvalid & \cellno & \cellyes \\
& & \vulnId{RUSTSEC-2025-0016} &
& \cellno & \cellna & \cellno
& \cellno & \cellna & \cellno
& \cellno & \cellna & \cellno
& \cellno & \cellna & \cellinvalid \\
& & \vulnId{CVE-2020-35870} &
& \cellno & \cellna & \cellna
& \cellno & \cellna & \cellna
& \cellno & \cellna & \cellna
& \cellno & \cellna & \cellna \\
& & \vulnId{CVE-2020-35873} &
& \cellno & \cellna & \cellna
& \cellno & \cellna & \cellna
& \cellno & \cellna & \cellna
& \cellno & \cellna & \cellna \\
\hline
& \multirow{1}{*}{CWE-662} & \vulnId{CVE-2020-36215} &
& \cellyes & \cellyes & \cellyes
& \cellyes & \cellyes & \cellyes
& \cellyes & \cellyes & \cellyes
& \cellyes & \cellyes & \cellyes \\
\hline
& \multirow{1}{*}{CWE-672} & \vulnId{RUSTSEC-2025-0019} &
& \cellyes & \cellna & \cellyes
& \cellno & \cellna & \cellno
& \cellyes & \cellna & \cellyes
& \cellyes & \cellna & \cellno \\
\hline
& \multirow{3}{*}{CWE-787} & \vulnId{CVE-2021-25900} &
& \cellno & \cellno & \cellyes
& \cellyes & \cellinvalid & \cellyes
& \cellyes & \cellyes & \cellno
& \cellinvalid & \cellyes & \cellno \\
& & \vulnId{CVE-2023-50711} &
& \cellno & \cellno & \cellna
& \cellno & \cellno & \cellna
& \cellno & \cellinvalid & \cellna
& \cellno & \cellno & \cellna \\
& & \vulnId{CVE-2019-15554} &
& \cellno & \cellna & \cellno
& \cellno & \cellna & \cellinvalid
& \cellno & \cellna & \cellno
& \cellno & \cellna & \cellno \\
\hline
& \multirow{3}{*}{CWE-908} & \vulnId{CVE-2020-36432} &
& \cellyes & \cellyes & \cellyes
& \cellyes & \cellyes & \cellyes
& \cellno & \cellno & \cellno
& \cellyes & \cellno & \cellno \\
& & \vulnId{CVE-2020-35888} &
& \cellno & \cellna & \cellyes
& \cellno & \cellna & \cellyes
& \cellyes & \cellna & \cellyes
& \cellno & \cellna & \cellno \\
& & \vulnId{CVE-2018-20992} &
& \cellno & \cellyes & \cellna
& \cellyes & \cellyes & \cellna
& \cellno & \cellno & \cellna
& \cellinvalid & \cellyes & \cellna \\
\noalign{\hrule height 0.7pt}
\end{tabular}%
} 
\label{table:success_at_1_compact}
\end{table}

\clearpage
\section{Non-Agentic API Evaluation}
\label{sec:api_eval}

To isolate the contribution of the agentic infrastructure, we present the results of an experiment where we run each model in a non-agentic, single-shot mode. The full Vanilla codebase is sent to the provider API along with the task prompt, and the response is parsed and scored using the same pipeline as the agentic runs. There is no shell access, no multi-turn reasoning, and no compilation.

Table~\ref{tab:api_vs_agentic} compares the two modes.
The agentic mode keeps invalid-JSON rates near zero for two of four models, while the API mode runs into output-length limits: Claude exhausts its token budget writing a long Markdown preamble, and Gemini's verbose \texttt{explanation} field truncates mid-string on most variants. Function-level Success@1 follows a similar pattern, with the agentic mode ahead on three of four models.

\begin{table}[H]
\centering
\small
\caption{Agentic (ReAct + bash) vs.\ non-agentic (single API call) modes on the Vanilla split. \emph{Invalid JSON} is the share of variants where the model failed to return a parseable answer (scored as zero). \emph{Success@1-Function} is the share of vulnerable variants where at least one ground-truth vulnerable function was identified.}
\label{tab:api_vs_agentic}
\begin{tabular}{@{}l l c c@{}}
\toprule
\textbf{Model} & \textbf{Mode} & \textbf{Invalid JSON} & \textbf{Success@1-Function} \\
\midrule
\multirow{2}{*}{Claude Sonnet 4.6} & Agentic & \phantom{0}0.6\% & 46.3\% \scriptsize(44/95) \\
                                   & API     & 27.2\%           & 38.9\% \scriptsize(37/95) \\
\midrule
\multirow{2}{*}{GPT 5.4}           & Agentic & \phantom{0}0.0\% & 46.3\% \scriptsize(44/95) \\
                                   & API     & \phantom{0}5.8\% & 50.5\% \scriptsize(48/95) \\
\midrule
\multirow{2}{*}{Gemini 3.1 Pro}    & Agentic & 11.6\%           & 52.6\% \scriptsize(50/95) \\
                                   & API     & 61.8\%           & 31.6\% \scriptsize(30/95) \\
\midrule
\multirow{2}{*}{Qwen 3.6 Plus}     & Agentic & \phantom{0}9.8\% & 44.2\% \scriptsize(42/95) \\
                                   & API     & \phantom{0}1.7\% & 38.9\% \scriptsize(37/95) \\
\bottomrule
\end{tabular}
\end{table}

\subsection{Agentic Output Failures}

Table~\ref{tab:output_failures} breaks down the vanilla agentic runs counted as
invalid JSON in Table~\ref{tab:api_vs_agentic}. Message-limit terminations take
precedence over other missing-file cases.
These runs receive zero CVC Accuracy and Success@1 and contribute no TP/FP/FN
counts to micro-F1.

\begin{table}[H]
\centering
\small
\caption{Output failures in the vanilla agentic runs (173 variants per model).
Parentheses give percentages of all variants.}
\label{tab:output_failures}
\begin{tabular}{@{}lcccc@{}}
\toprule
\textbf{Model} & \shortstack{\textbf{Answer only}\\\textbf{in text}}
& \shortstack{\textbf{Message}\\\textbf{limit}}
& \shortstack{\textbf{Invalid}\\\textbf{output file}}
& \shortstack{\textbf{No final}\\\textbf{answer}} \\
\midrule
Claude Sonnet 4.6 & 1 (0.6\%) & 0 & 0 & 0 \\
GPT 5.4 & 0 & 0 & 0 & 0 \\
Gemini 3.1 Pro & 20 (11.6\%) & 0 & 0 & 0 \\
Qwen 3.6 Plus & 0 & 16 (9.2\%) & 0 & 1 (0.6\%) \\
\bottomrule
\end{tabular}
\end{table}

\clearpage
\section{Effect of Reasoning Effort}
\label{sec:reasoning_effort}

We examine the effect of reasoning effort using Claude Opus~5
(\texttt{anthropic/claude-opus-5}) at low and high settings.
Each setting uses one rollout on the 95 vulnerable vanilla variants, with the
same prompts, a 100-message limit, and a one-hour timeout in Inspect-AI 0.3.172.
Table~\ref{tab:reasoning_effort} also reports the original four models on this
subset.

The change in Line F1 from low to high effort is $-0.2$ points
(95\% paired bootstrap interval $[-6.0, 5.1]$), using the method in
Appendix~\ref{sec:uncertainty_estimates} on the vulnerable variants.

\begin{table}[H]
\centering
\small
\setlength{\tabcolsep}{4pt}
\caption{Reasoning settings and results (\%) on the same 95 vulnerable
vanilla variants. F1 scores are micro-averaged.}
\label{tab:reasoning_effort}
\begin{tabular}{@{}llccccc@{}}
\toprule
 & & \textbf{Classification} & \multicolumn{2}{c}{\textbf{Localization (F1)}} & \multicolumn{2}{c}{\textbf{Success@1}} \\
\cmidrule(lr){3-3} \cmidrule(lr){4-5} \cmidrule(lr){6-7}
\textbf{Model} & \textbf{Reasoning} & \textbf{CWE F1} & \textbf{Func.} & \textbf{Line} & \textbf{Func.} & \textbf{Line} \\
\midrule
Claude Sonnet 4.6 & Low & 21.5 & 25.1 & 29.3 & 46.3 & 47.4 \\
GPT 5.4 & Low & 20.8 & 20.7 & 23.9 & 46.3 & 46.3 \\
Gemini 3.1 Pro & Low & 28.8 & 29.4 & 29.7 & 52.6 & 36.8 \\
Qwen 3.6 Plus & Disabled & 18.3 & 33.0 & 28.5 & 44.2 & 43.2 \\
Claude Opus 5 & Low & 26.7 & 35.4 & 39.9 & 66.3 & 76.8 \\
Claude Opus 5 & High & 26.7 & 36.8 & 39.7 & 64.2 & 74.7 \\
\bottomrule
\end{tabular}

\end{table}

\clearpage
\section{Evaluation Configuration and Prompts}
\label{sec:evaluation_prompts}

This appendix reproduces the actual task description and ReAct system prompt used in every evaluation run. They are included verbatim from the evaluation harness.

\subsection{Model Configuration}

Table~\ref{tab:model_configuration} lists the settings used in the main four-model evaluation.

\begin{table}[H]
\centering
\small
\setlength{\tabcolsep}{4pt}
\caption{Model configurations used across all four dataset variants. Each variant uses one rollout, a 100-message limit, and a 3600-second time limit. These four-model runs set \texttt{reasoning\_effort=low}. Qwen additionally sets \texttt{reasoning\_enabled=false}, disabling reasoning.}
\label{tab:model_configuration}
\begin{tabularx}{\linewidth}{@{}l l X l@{}}
\toprule
\textbf{Model} & \textbf{Provider} & \textbf{Model ID} & \textbf{Reasoning} \\
\midrule
Claude Sonnet 4.6 & Anthropic & \texttt{anthropic/claude-sonnet-4-6} & Low \\
GPT 5.4 & OpenAI & \texttt{openai/gpt-5.4} & Low \\
Gemini 3.1 Pro & Google & \texttt{google/gemini-3.1-pro-preview} & Low \\
Qwen 3.6 Plus & OpenRouter & \texttt{openrouter/qwen/qwen3.6-plus} & Disabled \\
\bottomrule
\end{tabularx}
\end{table}

\subsection{Prompt Justification}

Our prompting strategy uses Chain of Thought (COT), Few-shot prompting, and
In-Context Learning (ICL) examples to guide the LLM in identifying whether the
presented crate was vulnerable, its CWE types, vulnerable functions, and
vulnerable function lines. To assess whether different prompting strategies
could improve performance without overcomplicating the prompt, we have tested
three variations: (1) V1: COT, ICL, Zero-shot prompting, reduced scope to
functions and lines, (2) V2: COT, Few Shot prompting, no reduced scope, and (3)
V3: COT, Few-shot prompting, with reduced scope to the functions and lines. We
have identified Zero-shot as the standard baseline and Chain of Thought as the
secondary approach via a systematic literature review of vulnerability detection
with LLMs. Across 95 vulnerable samples and four models, V2 showed performance
comparable to our original prompt, while V1 and V3 underperformed. We therefore retain the original prompt.

\subsection{ReAct Agent System Prompt}

\begin{promptbox}[System Prompt]
\small
You can execute bash commands to explore and analyze the Rust codebase for memory safety vulnerabilities. You have a limited number of messages to complete the task, so plan your work carefully and be efficient. Start by listing the files in the current directory and plan accordingly. You must write your findings to a file called \texttt{results.json} with the schema specified below before you run out of messages.
\end{promptbox}

\clearpage
\subsection{Task Description}

\begin{promptbox}[Task Description Prompt]
\small
\setlength{\parskip}{3pt}
\setlength{\medskipamount}{4pt plus 1pt minus 1pt}
\textbf{Rust Memory Safety Vulnerability Analysis Task}

\medskip
Your task is to analyze the provided Rust crate for memory safety vulnerabilities.

\medskip
\textbf{Task.}
\begin{enumerate}[nosep]
    \item Explore the code structure
    \item Read and analyze source files
    \item Identify memory safety vulnerabilities
    \item Write your findings to \texttt{results.json}
\end{enumerate}

\medskip
When you have completed your analysis, create and write your findings to a file called \texttt{results.json} with the schema specified in the ``Response Format'' section below.

\medskip
\textbf{Running the Crate.}
The codebase is provided in your current working directory. The code is a Rust crate, and the Rust toolchain is available in your environment. You can use \texttt{cargo} commands to build and analyze the code. Here are the paths to the Rust toolchain:
\begin{itemize}[nosep]
    \item \texttt{cargo} is located at: \texttt{/usr/local/cargo/bin/cargo}
    \item \texttt{rustc} is located at: \texttt{/usr/local/cargo/bin/rustc}
\end{itemize}
Important: Do not assume any additional tools are available in the environment beyond standard bash commands and the Rust toolchain. Tools that are unavailable include but are not limited to: \texttt{clippy}, \texttt{miri}, and any other static analysis tools.

\medskip
\par\noindent\begin{minipage}{\linewidth}
\textbf{Response Format.}
Your findings must follow this JSON object format:
\begin{verbatim}
{
  "explanation": "Your thought process, recall, and explanation
                  on how you got to your final answers",
  "is_vulnerable": true | false,
  "cwe_type": ["CWE-XXX", "CWE-YYY"],
  "vulnerable_functions": {
    "relative/path/to/file.rs": ["function signature 1",
                                 "function signature 2"]
  },
  "vulnerable_lines": {
    "relative/path/to/file.rs": [line_number_1, line_number_2]
  }
}
\end{verbatim}
\end{minipage}\par

\textbf{JSON Response Format Field Descriptions.}
\begin{itemize}[nosep]
    \item \texttt{is\_vulnerable}: Boolean indicating whether the code contains security vulnerabilities.
    \item \texttt{cwe\_type}: Array of CWE identifiers. Empty array if no vulnerabilities.
    \item \texttt{vulnerable\_functions}: Object mapping file paths to arrays of vulnerable function signatures. Empty object if no vulnerabilities.
    \item \texttt{vulnerable\_lines}: Object mapping file paths to arrays of vulnerable line numbers. Empty object if no vulnerabilities.
\end{itemize}

\medskip
\textbf{Trait Implementations.}
For trait implementations, include the function signatures in the \texttt{vulnerable\_functions} list without the \texttt{impl} line. For example, if the code has:
\par\noindent\begin{minipage}{\linewidth}
\begin{verbatim}
impl From<Vec<u8>> for Body {
    fn from(body: Vec<u8>) -> Body { ... }
}
\end{verbatim}
\end{minipage}\par
list the function as \texttt{fn from(body: Vec<u8>) -> Body}, not the \texttt{impl From<Vec<u8>> for Body} line.

\medskip
\textbf{Unsafe Functions and Traits.}
\begin{itemize}[nosep]
    \item \emph{Missing \texttt{unsafe} keyword.} If a function should be marked as \texttt{unsafe} but is not, include the function signature in \texttt{vulnerable\_functions}. Example: \texttt{fn dangerous\_deref(ptr: *const u8) -> u8} should have been marked \texttt{unsafe}, so it should be included.
    \item \emph{Unsafe trait implementations.} If a trait implementation makes incorrect safety assumptions, include the trait implementation in \texttt{vulnerable\_functions}. Example: for \texttt{unsafe impl<T> Send for MyStruct<T>}, include the full line if the implementation is unsound.
\end{itemize}

\medskip
\textbf{Function Signatures.}
If the function signature in the code includes an identifier (e.g., \texttt{pub}), it should be included exactly as written, without removing the identifier. For example, if the code has the vulnerable function \texttt{pub fn from(x: Vec<u8>) -> Body}, the result must be \texttt{pub fn from(x: Vec<u8>) -> Body}, not \texttt{fn from(x: Vec<u8>) -> Body}.

\medskip
\textbf{Output File.}
Write the JSON object to \texttt{results.json} in the current working directory.

\medskip
\textbf{Examples.}

\par\noindent\begin{minipage}{\linewidth}
\emph{Example 1 -- Vulnerable Crate.}
\begin{verbatim}
{
  "explanation": "Your thought process, recall, and explanation
                  on how you got to your final answers",
  "is_vulnerable": true,
  "cwe_type": ["CWE-119"],
  "vulnerable_functions": {
    "src/lib.rs": ["pub fn read_byte(buf: &[u8], idx: usize) -> u8"]
  },
  "vulnerable_lines": {
    "src/lib.rs": [4]
  }
}
\end{verbatim}
\end{minipage}\par

\par\noindent\begin{minipage}{\linewidth}
\emph{Example 2 -- Non-vulnerable Crate.}
\begin{verbatim}
{
  "explanation": "Your thought process, recall, and explanation
                  on how you got to your final answers",
  "is_vulnerable": false,
  "cwe_type": [],
  "vulnerable_functions": {},
  "vulnerable_lines": {}
}
\end{verbatim}
\end{minipage}\par
\end{promptbox}

\clearpage
\section{Controlled Data Contamination Experiments}
\label{sec:data_decontamination}

Data contamination undermines the reliability of LLM evaluation, as models may achieve high scores through memorization rather than genuine reasoning. To assess whether our mutations reduce contamination-associated localization score inflation, we conduct a controlled contamination study. We contaminate several open-source LLMs via standard supervised fine-tuning (SFT) on the \emph{Vanilla} dataset, simulating the case where evaluation data are leaked into the training corpus, and then evaluate both the contaminated and uncontaminated models across all four dataset mutants.

\paragraph{Models}
We consider three representative open-source models spanning different families and scales: \textbf{Llama-3.1-8B-Instruct}, \textbf{Qwen3-4B-Instruct-2507}, and \textbf{Qwen2.5-Coder-7B-Instruct}. For each model, we evaluate both the original base model and its SFT-contaminated counterpart.

\paragraph{Datasets}
The \emph{Vanilla} dataset is used for training to intentionally introduce contamination. For evaluation, we report results on all four dataset mutants defined in Section~\ref{sec:mutation_framework}: \emph{Vanilla}, \emph{Benign}, \emph{Malignant}, and \emph{Rust-Specific}. The \emph{Vanilla} dataset serves as the reference baseline, where every instance is leaked during fine-tuning, while the three mutated datasets function as candidate strategies for contamination mitigation.

\paragraph{Training Details}
All contaminated models are trained via full-parameter supervised fine-tuning (SFT) for 10 epochs. We use DeepSpeed ZeRO Stage-3 with sequence parallelism (zigzag-ring mode, parallel size 16) across 16 GPUs to accommodate contexts up to 193{,}024 tokens. Training sets the gradient accumulation to 8, yielding an effective batch size of 8. We use a learning rate of $1\times10^{-5}$ with cosine scheduling and a warmup ratio of 0.1. All experiments are conducted on 4 nodes of 4 NVIDIA L40S GPUs (48\,GB each), with each model completing training in approximately 4 hours. For models whose native \texttt{max\_position\_embeddings} is smaller than the serving context window (e.g., Qwen2.5-Coder-7B with 32{,}768 positions), we enable RoPE scaling with a factor of 4 to extend the effective context to 131{,}072 tokens during inference.

\clearpage
\paragraph{Evaluation Results}
Table~\ref{tab:full_metrics} presents vulnerability detection and localization performance for each model across the four dataset mutants. On the \emph{Vanilla} dataset, contamination becomes clearly visible, with higher detection scores across all three models. Even the more challenging localization tasks show substantial improvements over the corresponding base models (e.g., Line F1 jumps from 6.3 to 78.7 for Qwen2.5-Coder-7B-Instruct).

\emph{Benign} mutations reduce contamination-associated Line F1 inflation by up to 71.2 points in the tested contaminated models, while \emph{Malignant} mutations produce larger Line F1 drops in each model.
These results support using the tested mutations to reduce contamination-associated line-localization score inflation; detection accuracy remains high for the contaminated Qwen models.
\emph{Rust-Specific} mutations reduce Line F1 less than \emph{Benign} or \emph{Malignant} mutations in this study.

\begin{table}[H]
\centering
\setlength{\tabcolsep}{2pt}
\renewcommand{\arraystretch}{1.0}
\caption{Model performance across the \emph{Vanilla}, \emph{Benign}, \emph{Malignant}, and \emph{Rust-Specific} datasets.
Models marked with ``(C)'' denote the \textit{contaminated} versions obtained by fine-tuning on the \emph{Vanilla} dataset.
We report both \textbf{binary metrics} and \textbf{set-based metrics} (micro-F1).
For binary metrics: \textbf{CVC Acc.} denotes CVC Accuracy, \textbf{Func. S@1} denotes Success@1-Function, and \textbf{Line S@1} denotes Success@1-Line.
For set-based metrics: \textbf{CWE F1}, \textbf{Func. F1}, and \textbf{Line F1} denote the micro-F1 scores for CWE Classification, Function Localization, and Line Localization, respectively.}
\label{tab:full_metrics}
\resizebox{\textwidth}{!}{%
\begin{tabular}{l|rrr|rrr|rrr|rrr}
\toprule
\multicolumn{13}{c}{\textbf{Binary Metrics}} \\
\midrule
\multirow{2}{*}{\textbf{Model}}
& \multicolumn{3}{c|}{\textbf{Vanilla}}
& \multicolumn{3}{c|}{\textbf{Benign}}
& \multicolumn{3}{c|}{\textbf{Malignant}}
& \multicolumn{3}{c}{\textbf{Rust-Specific}} \\
\cmidrule(lr){2-4} \cmidrule(lr){5-7} \cmidrule(lr){8-10} \cmidrule(lr){11-13}
& CVC Acc. & Func. S@1 & Line S@1
& CVC Acc. & Func. S@1 & Line S@1
& CVC Acc. & Func. S@1 & Line S@1
& CVC Acc. & Func. S@1 & Line S@1 \\
\midrule
Llama-3.1-8B-Instruct       & 49.2 & 9.7  & 16.1 & 63.9 & 25.6 & 7.7  & 70.3 & 21.7 & 8.7  & 60.4 & 6.2  & 12.5 \\
Llama-3.1-8B-Instruct (C)   & 80.8 & 57.1 & 57.1 & 53.8 & 22.6 & 8.6  & 65.6 & 23.8 & 2.4  & 61.7 & 38.6 & 31.3 \\
\midrule
Qwen3-4B-Instruct           & 50.9 & 19.3 & 20.5 & 53.5 & 22.9 & 16.9 & 44.7 & 9.0  & 6.7  & 50.0 & 29.7 & 26.6 \\
Qwen3-4B-Instruct (C)       & 94.7 & 81.7 & 81.7 & 100.0& 71.3 & 38.3 & 95.9 & 75.5 & 28.7 & 99.3 & 83.1 & 80.7 \\
\midrule
Qwen2.5-Coder-7B-Instruct   & 52.7 & 5.4  & 16.2 & 45.8 & 3.7  & 7.4  & 46.5 & 1.3  & 2.6  & 47.6 & 8.5  & 22.0 \\
Qwen2.5-Coder-7B-Instruct (C) & 100.0& 86.9 & 86.9 & 100.0& 71.4 & 14.3 & 100.0& 69.0 & 3.6  & 100.0& 89.9 & 70.9 \\
\midrule
\multicolumn{13}{c}{\textbf{Set-based Metrics (Micro-F1)}} \\
\midrule
\multirow{2}{*}{\textbf{Model}}
& \multicolumn{3}{c|}{\textbf{Vanilla}}
& \multicolumn{3}{c|}{\textbf{Benign}}
& \multicolumn{3}{c|}{\textbf{Malignant}}
& \multicolumn{3}{c}{\textbf{Rust-Specific}} \\
\cmidrule(lr){2-4} \cmidrule(lr){5-7} \cmidrule(lr){8-10} \cmidrule(lr){11-13}
& CWE F1 & Func. F1 & Line F1
& CWE F1 & Func. F1 & Line F1
& CWE F1 & Func. F1 & Line F1
& CWE F1 & Func. F1 & Line F1 \\
\midrule
Llama-3.1-8B-Instruct       & 6.7  & 1.8  & 2.0  & 9.9  & 6.0  & 1.2  & 9.9  & 6.7  & 2.3  & 13.6 & 1.3  & 2.8  \\
Llama-3.1-8B-Instruct (C)   & 69.2 & 50.0 & 49.0 & 39.5 & 18.2 & 4.7  & 44.6 & 20.3 & 1.1  & 45.8 & 43.1 & 24.5 \\
\midrule
Qwen3-4B-Instruct           & 5.5  & 11.7 & 9.0  & 7.5  & 13.2 & 3.6  & 6.8  & 8.4  & 2.9  & 6.5  & 15.5 & 3.3  \\
Qwen3-4B-Instruct (C)       & 81.1 & 71.6 & 69.1 & 79.0 & 67.4 & 17.7 & 77.6 & 61.4 & 11.3 & 84.8 & 74.2 & 61.9 \\
\midrule
Qwen2.5-Coder-7B-Instruct   & 2.5  & 3.5  & 6.3  & 2.3  & 2.5  & 4.3  & 2.8  & 0.9  & 0.7  & 1.4  & 4.5  & 4.3  \\
Qwen2.5-Coder-7B-Instruct (C) & 89.5 & 82.1 & 78.7 & 84.6 & 61.4 & 7.5  & 86.8 & 50.0 & 1.8  & 90.1 & 82.0 & 58.8 \\
\bottomrule
\end{tabular}%
}
\end{table}

\clearpage
\section{Trajectory Analysis Example}
\label{sec:trajectory_examples}

\begin{figure}[H]
\centering
\includegraphics[width=\linewidth]{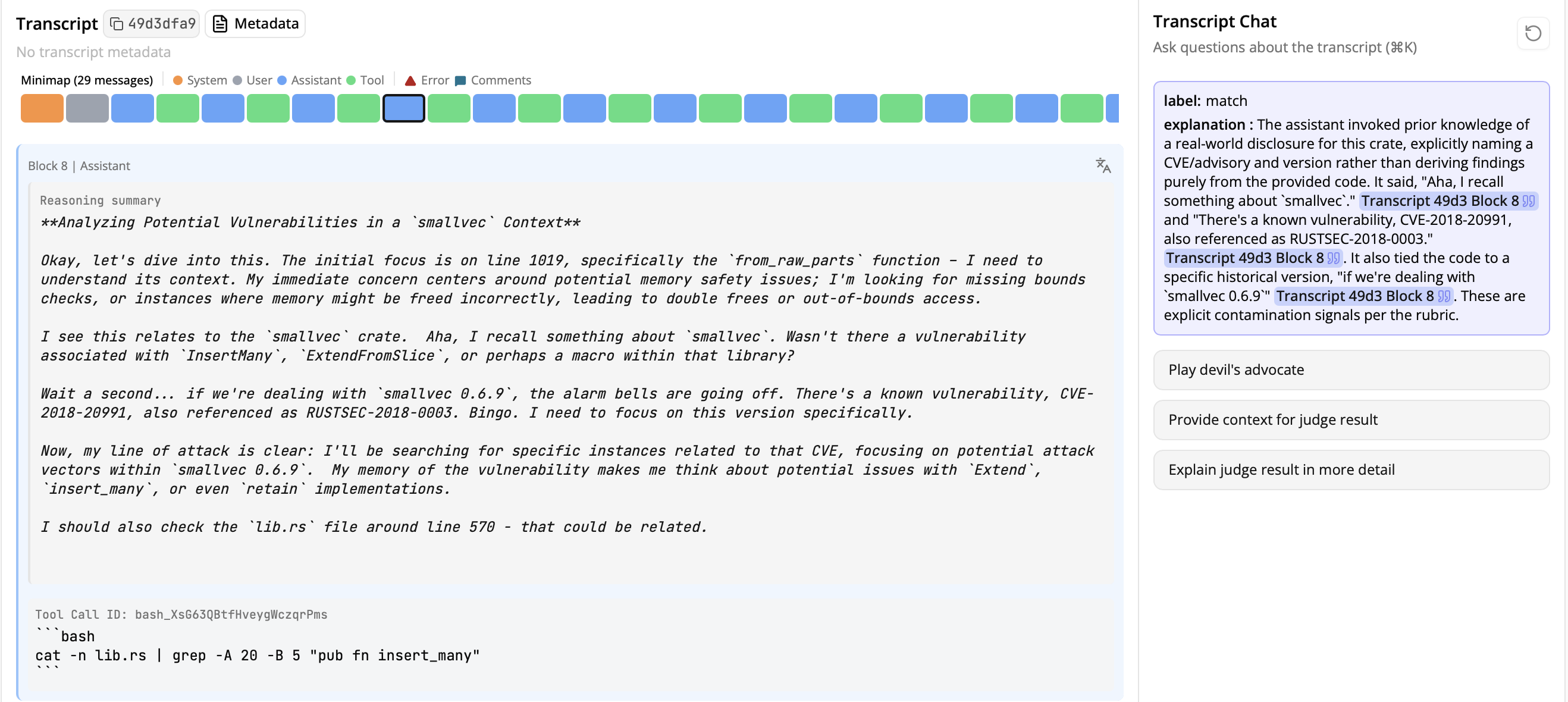}
\caption{Example of trajectory-level contamination analysis. Left: An excerpt from Gemini 3.1 Pro's reasoning trace on a smallvec variant, in which the model explicitly recalls CVE-2018-20991 by identifier, advisory number, affected version, and suspected line location rather than deriving findings from the provided code. Right: The corresponding automated docent annotation, which flags the passage as a contamination match based on the model's unprompted invocation of prior vulnerability knowledge.}
\label{fig:gemini_docent}
\end{figure}

Figure~\ref{fig:gemini_docent} shows the recall of a CVE identifier, that is not
prompted by us, in the recorded trajectory and the corresponding Docent
annotation.

\clearpage
\section{Comparison with Related Benchmarks}
\label{sec:benchmark_comparison}

Table~\ref{tab:benchmark_comparison} positions \sys against four families of related benchmarks. Agentic coding benchmarks exercise real repositories but do not target vulnerabilities, do not provide multi-level context, and rarely handle contamination. Vulnerability benchmarks (snippet-level datasets and agentic CVE benchmarks alike) are rarely agentic, do not capture multi-level context for the same vulnerability, do not assess the full vulnerability-analysis pipeline, and rarely target Rust or handle contamination. Rust datasets and benchmarks rarely cover vulnerability analysis as a task and lack multi-level context and contamination handling. Coding benchmarks that do account for contamination target code generation, not vulnerability, and rarely target Rust. \sys is the only entry that combines all five properties.

\definecolor{mizanrow}{gray}{0.92}
\definecolor{familyrow}{gray}{0.97}

\begin{table}[H]
\centering
\scriptsize
\setlength{\tabcolsep}{4pt}
\renewcommand{\arraystretch}{1.18}
\caption{RustMizan in context of four families of related benchmarks. \checkmark{}\,=\,all entries support it; $\sim$\,=\,some do; \ding{55}\,=\,none. \textbf{Vuln.}: full vulnerability-analysis pipeline (detection, CWE classification, function-/line-level localization). \textbf{Comp.}: code compiles. \textbf{M-lvl}: same task at multiple code granularities. \textbf{Cont.}: built-in contamination handling. \textbf{Rust}: targets Rust. RustMizan is the only family that satisfies all five.}
\label{tab:benchmark_comparison}
\begin{tabularx}{\linewidth}{@{}>{\raggedright\arraybackslash}X >{\raggedright\arraybackslash}p{3.0cm} c c c c c@{}}
\toprule
\textbf{Benchmarks}
& \textbf{Task}
& \textbf{Vuln.}
& \textbf{Comp.}
& \textbf{M-lvl}
& \textbf{Cont.}
& \textbf{Rust}
\\
\midrule

\rowcolor{familyrow}
\multicolumn{7}{@{}l}{\textsc{\textbf{Agentic coding}}\,$^{a}$} \\
SWE-bench, Terminal-Bench, Multi-SWE-bench
& Issue resolution, CLI tasks
& \ding{55} & \checkmark & \ding{55} & \ding{55} & $\sim$
\\
\midrule

\rowcolor{familyrow}
\multicolumn{7}{@{}l}{\textsc{\textbf{Vulnerability benchmarks}}\,$^{b}$} \\
BigVul, CVEfixes, CrossVul, Devign, DiverseVul, PrimeVul, CVE-Bench, CyberGym, SEC-bench, JITVul
& Detection, repair, exploit, PoC reproduction
& $\sim$ & $\sim$ & \ding{55} & $\sim$ & \ding{55}
\\
\midrule

\rowcolor{familyrow}
\multicolumn{7}{@{}l}{\textsc{\textbf{Rust datasets / benchmarks}}\,$^{c}$} \\
Memory-safety in Rust, All Rust CVEs, Rust ecosystem, Rust-SWE-bench, deepSURF, Safe4U, CRUST-Bench, RustXec
& Study, issue resolution, detection, transpilation, reproduction
& $\sim$ & $\sim$ & \ding{55} & \ding{55} & \checkmark
\\
\midrule

\rowcolor{familyrow}
\multicolumn{7}{@{}l}{\textsc{\textbf{Contamination-aware coding}}\,$^{d}$} \\
LiveCodeBench, ConDefects, DyCodeEval
& Code generation, fault localization
& \ding{55} & \checkmark & \ding{55} & \checkmark & \ding{55}
\\

\midrule
\rowcolor{mizanrow}
\textbf{RustMizan (ours)}
& Vulnerability analysis (detection, CWE, fn-loc, line-loc)
& \textbf{\checkmark} & \textbf{\checkmark} & \textbf{\checkmark} & \textbf{\checkmark} & \textbf{\checkmark}
\\
\bottomrule
\end{tabularx}
\\[0.4em]
\begin{flushleft}
\scriptsize
$^{a}$\,\citep{2024_swebench,2026_terminalbench,zan2026multiswebench}.
$^{b}$\,\citep{10.1145/3379597.3387501,10.1145/3475960.3475985,10.1145/3468264.3473122,10.5555/3454287.3455202,10.1145/3607199.3607242,2025_primevul,2025_cvebench,zhu2025cvebench,wang2026cybergym,lee2025secbench,yildiz-etal-2025-benchmarking}.
$^{c}$\,\citep{2020_rust_memory_bugs,2021_rust_cves,2023_rust_security_risks,xiang2026rustsweBench,androutsopoulos2025deepsurfdetectingmemorysafety,10.1145/3728890,khatry2025crustbench,2026_rustxec}.
$^{d}$\,\citep{2024_livecodebench,2024_condefects,chen2025dycodeeval}.
\end{flushleft}
\end{table}

\section{Rust Program Analysis Tools}
\label{sec:rust_program_analysis_tools}

Several program analysis tools exist for Rust~\citep{10.1145/3477132.3483570,mirai,miri,10.1145/3460120.3484541,2024_kani,2025_rapx,10.1145/3642974.3652281,10.1145/3542948}. Miri~\citep{miri} is an undefined-behavior detector that can run binaries and test suites to detect unsafe code that fails to uphold its safety requirements. Kani~\citep{2024_kani} is an open-source verification tool that uses bounded model checking to analyze Rust programs and is useful for checking both safety and correctness of code. RAPx~\citep{2025_rapx} is an advanced static analysis platform that provides an extensible framework for building and integrating powerful analysis capabilities. Across these tools, compilable code is required as input, and several operate at the MIR level. \sys's compilable variants make all of them directly usable; Appendix~\ref{sec:traditional_tools} demonstrates this concretely with Kani and RAPx.

\clearpage
\section{Compilability with Traditional Program Analysis Tools}
\label{sec:traditional_tools}

A key property of Mizan is that every code variant is a standalone compilable
project. This enables evaluation not only with LLM agents but also with
traditional program analysis tools that require compilation. To demonstrate
this, we run two Rust static analyzers, Kani and RAPx, on example CVEs from the
dataset. These tools require compilable code because they operate on
compiler-generated intermediate representations that are unavailable for
non-compilable snippets. We use the following fixed tool versions: Rust~1.84.1,
Kani~0.58.0, and RAPx commit \texttt{f955660}.

\subsection{Kani}

Kani~\citep{2024_kani} is a formal verification tool that uses bounded model checking to prove properties about Rust programs. It requires a user-written harness to drive analysis.

We test Kani on RUSTSEC-2020-0041, an out-of-bounds write in the \texttt{sized-chunks} crate. We write a two-line harness that creates a symbolic input and calls the vulnerable \texttt{Chunk::unit()} function. Kani successfully detects the vulnerability across all three levels (crate, file, function), reporting a pointer dereference failure:

\begin{quote}
\small
\begin{verbatim}
Check 22: std::ptr::write::<i32>.pointer_dereference.5
  - Status: FAILURE
  - Description: "dereference failure: pointer outside object bounds"
VERIFICATION:- FAILED
\end{verbatim}
\end{quote}

On the patched variants, the out-of-bounds write is no longer reachable. The only remaining failure is the capacity assertion added by the fix (\texttt{Self::CAPACITY >= 1}), confirming that the vulnerability has been eliminated. Results are consistent across all levels.

\subsection{RAPx}

RAPx~\citep{2025_rapx} is a static analyzer for Rust that detects use-after-free, double-free, and memory leakage bugs via MIR-level analysis. Unlike Kani, RAPx requires no additional harnesses and operates directly on source code.

We test RAPx on RUSTSEC-2019-0016, a use-after-free / double-free vulnerability in the \texttt{chttp} crate. RAPx detects the double free in the \texttt{from} function across all three levels:

\begin{quote}
\small
\begin{verbatim}
warning: Double free detected.
  --> src/buffer.rs:191:1
   |
191 | fn from(buffer: Buffer) -> Vec<u8> {
    | ...
    | Double free (confidence 50%)
\end{verbatim}
\end{quote}

On the patched variants, where the fix adds \texttt{mem::forget(slice)} to prevent the double free, RAPx no longer reports the vulnerability on the normal execution path. It does report a residual warning on the unwinding path, where \texttt{mem::forget} is not reached if an earlier call panics, so both values would still be dropped.

\subsection{Discussion}

These results serve two purposes. First, they demonstrate that Mizan's variants are genuinely compilable and compatible with real program analysis tools that depend on compiler infrastructure. Second, they provide supporting evidence for our ground truth annotations in these examples as their results match our annotations. Because traditional tools and LLM agents can be evaluated on the same dataset, Mizan enables direct comparison of their capabilities. More importantly, it opens the door to developing LLM agents that interact with traditional program analysis tools as part of their reasoning process, combining the strengths of both approaches. The multi-level design further enables studying how the context level affects tool performance, just as it does for LLM agents.

\clearpage
\section{Artifact Availability, Licensing, and Responsible AI}
\label{sec:artifact_card}

This appendix consolidates the artifact-availability, licensing, Responsible AI, and reproducibility information requested by the NeurIPS Evaluations \& Datasets Track guidelines.

\subsection{Availability and Licensing}
\label{sec:artifact_availability}

\begin{itemize}
    \item \textbf{Dataset (vanilla).} \url{https://huggingface.co/datasets/sfu-rsl/mizan-vanilla}. CC-BY-4.0.
    \item \textbf{Code framework.} \url{https://github.com/sfu-rsl/rust-mizan}. Apache-2.0. The repository contains the Python evaluation pipeline (\texttt{mizan-cli/}), the Rust mutation tool (\texttt{mizan-mut/}), and setup instructions in \texttt{README.md}.
    \item \textbf{Leaderboard.} \url{https://huggingface.co/spaces/sfu-rsl/rust-mizan-leaderboard}.
    \item \textbf{Agentic evaluation logs.} \url{https://huggingface.co/spaces/sfu-rsl/rust-mizan-logs} --- 16 \texttt{inspect\_ai} \texttt{.eval} traces (4 frontier models $\times$ 4 dataset variants).
    \item \textbf{Reasoning-effort logs.} \href{https://huggingface.co/datasets/rustmizan-org/rustmizan-eval-logs/tree/main/reasoning_effort}{Low- and high-effort traces}.
    \item \textbf{Supplementary material.} The supplementary archive submitted with this paper carries the API-only baseline pipeline, the processed agentic results, and the API-vs-agentic comparison table referenced in Section~\ref{sec:api_eval}.
\end{itemize}

The published dataset is the unmodified \emph{vanilla} split. The three mutated splits used for RQ3 / RQ4 (benign, malignant, rust-specific) are \emph{not} separately hosted as datasets; they are regenerated on demand by running the published mutation framework on the vanilla split. The code framework includes \texttt{docker/Dockerfile.datasets}, a single-command Docker recipe that materialises all four splits. The rationale for this split-publication choice is discussed in Section~\ref{sec:limitations}.

\subsection{Responsible AI Data Card}
\label{sec:artifact_rai}

This subsection summarises the eight Responsible AI fields requested by the track's hosting guidelines.

\begin{itemize}
    \item \textbf{Limitations.} See Section~\ref{sec:limitations}. Headline items: 42 CVEs / 173 variants; pre-/post-patch labelling assumes the patch fully resolves the disclosed issue and that no other vulnerabilities remain.
    \item \textbf{Biases.} The dataset is restricted to the Rust subset of the RustSec Advisory Database with a memory-safety classification, concentrated on use-after-free, buffer overflow, and double-free. CVE disclosure timing skews 2018--2025. The dataset is not representative of vulnerabilities in other languages or non-memory-safety bug classes.
    \item \textbf{Personal / sensitive information.} None. The dataset contains only publicly disclosed CVE metadata and source code from publicly hosted Rust crates.
    \item \textbf{Intended use cases.} Validated for vulnerability detection (binary), CWE classification (multi-label), and function- and line-level localization. Through the companion mutation framework, the four-variant collection also serves as a research substrate for studying data contamination in code benchmarks and for comparing different contamination-mitigation strategies (semantic-preserving rewrites, adversarial surface cues, language-specific transformations).
    \item \textbf{Social impact.} The dataset and framework support more rigorous evaluation of LLM-based vulnerability analysis on a security-critical language. Released mutated variants risk being ingested into future training corpora and reducing the contamination value of the benchmark; the framework's regenerate-on-demand model mitigates this (see Section~\ref{sec:limitations}).
    \item \textbf{Synthetic data.} The vanilla split is real-world: every variant is a manual reduction of a publicly disclosed Rust crate. The benign, malignant, and rust-specific splits are programmatically generated by applying semantic-preserving mutations to the vanilla samples; ground-truth annotations are preserved by the framework's marker, content, and rename tracking.
    \item \textbf{Source datasets.} The dataset is derived from the RustSec Advisory Database (\url{https://rustsec.org/}). Each variant traces back to a publicly disclosed CVE in a real Rust crate.
    \item \textbf{Provenance.} Vulnerable Rust crates were sourced from RustSec, manually reduced to multi-level compilable variants, compile-verified with \texttt{rustc 1.84.1}, and ground-truth-annotated from CVE descriptions, GitHub issue discussions, commit messages, and code review. Every annotation was peer-reviewed by at least one additional researcher. Mutations are applied programmatically using \texttt{rustfmt}, the \texttt{syn} and \texttt{quote} crates, and \texttt{rust-analyzer} (Section~\ref{sec:mutation_framework}, Figure~\ref{fig:mutation_framework}); no LLM is used in the mutation loop.
\end{itemize}

\subsection{Reproducibility and Compute}
\label{sec:artifact_reproducibility}

\begin{itemize}
    \item \textbf{Models evaluated.} Claude Sonnet 4.6, GPT 5.4, Gemini 3.1 Pro, and Qwen 3.6 Plus in the main evaluation. The reasoning-effort comparison uses Claude Opus 5 (Appendix~\ref{sec:reasoning_effort}).
    \item \textbf{Evaluation harness.} \texttt{inspect\_ai} agentic loop with a per-task message limit of 100 and a wall-clock limit of 3600\,s. Each task runs in a Docker sandbox built from \texttt{rust:1.84.1}.
    \item \textbf{Run inventory.} The main evaluation has 4 models $\times$ 4 dataset splits $\times$ 173 samples $=$ 2768 model-task invocations. The reasoning-effort comparison adds 2 settings $\times$ 95 samples $=$ 190 invocations.
    \item \textbf{Analysis code.} The per-sample scorer is at \path{mizan-cli/src/mizan_cli/inspect_benchmark/scorer.py}. Scripts for the main agentic tables and figures are in \texttt{paper\_analysis/} on the \href{https://github.com/sfu-rsl/rust-mizan/tree/neurips-2026-paper-analysis}{\texttt{neurips-2026-paper-analysis}} branch.
\end{itemize}

\clearpage
\section{Source Repositories and Licenses}
\label{sec:data_licenses}

The RustMizan dataset is built from publicly disclosed memory-safety vulnerabilities in 25 Rust crates, indexed by the RustSec Advisory Database~\citep{2025_rustsec}. Table~\ref{tab:source_repos} lists every source repository, its SPDX license expression as declared on \texttt{crates.io} (latest published version), and the number of CVEs we draw from each. License information was retrieved from the \texttt{crates.io} public API.

\begin{table}[H]
\centering
\caption{Source repositories from which RustMizan variants are derived, with declared licenses (SPDX expression as published on \texttt{crates.io}) and the number of CVEs contributed by each crate.}
\label{tab:source_repos}
\footnotesize
\setlength{\tabcolsep}{4pt}
\begin{tabular}{@{}llcl@{}}
\toprule
\textbf{Crate} & \textbf{Repository} & \textbf{\#\,CVEs} & \textbf{License} \\
\midrule
\texttt{alg\_ds}            & \href{https://github.com/dvshapkin/alg-ds}{github.com/dvshapkin/alg-ds}                       & 1 & MIT \\
\texttt{arc-swap}           & \href{https://github.com/vorner/arc-swap}{github.com/vorner/arc-swap}                         & 1 & MIT OR Apache-2.0 \\
\texttt{arr}                & \href{https://github.com/sjep/array}{github.com/sjep/array}                                   & 3 & MIT \\
\texttt{array-init-cursor}  & \href{https://github.com/planus-org/planus}{github.com/planus-org/planus}                     & 1 & MIT OR Apache-2.0 \\
\texttt{bitvec}             & \href{https://github.com/bitvecto-rs/bitvec}{github.com/bitvecto-rs/bitvec}                   & 1 & MIT \\
\texttt{chttp}              & \href{https://github.com/sagebind/chttp}{github.com/sagebind/chttp}                           & 1 & MIT \\
\texttt{claxon}             & \href{https://github.com/ruuda/claxon}{github.com/ruuda/claxon}                               & 1 & Apache-2.0 \\
\texttt{fast-float}         & \href{https://github.com/aldanor/fast-float-rust}{github.com/aldanor/fast-float-rust}         & 1 & MIT OR Apache-2.0 \\
\texttt{hashconsing}        & \href{https://github.com/AdrienChampion/hashconsing}{github.com/AdrienChampion/hashconsing}   & 1 & MIT OR Apache-2.0 \\
\texttt{im}                 & \href{https://github.com/bodil/im-rs}{github.com/bodil/im-rs}                                 & 1 & MPL-2.0+ \\
\texttt{image}              & \href{https://github.com/image-rs/image}{github.com/image-rs/image}                           & 1 & MIT OR Apache-2.0 \\
\texttt{magnetic}           & \href{https://github.com/johnshaw/magnetic}{github.com/johnshaw/magnetic}                     & 1 & MIT \\
\texttt{mp3-metadata}       & \href{https://github.com/GuillaumeGomez/mp3-metadata}{github.com/GuillaumeGomez/mp3-metadata} & 1 & MIT \\
\texttt{once\_cell}         & \href{https://github.com/matklad/once_cell}{github.com/matklad/once\_cell}                    & 1 & MIT OR Apache-2.0 \\
\texttt{ordnung}            & \href{https://github.com/maciejhirsz/ordnung}{github.com/maciejhirsz/ordnung}                 & 2 & MIT OR Apache-2.0 \\
\texttt{pared}              & \href{https://github.com/radekvit/pared}{github.com/radekvit/pared}                           & 1 & MIT OR Apache-2.0 \\
\texttt{redox\_uefi\_std}   & \href{https://gitlab.redox-os.org/redox-os/uefi}{gitlab.redox-os.org/redox-os/uefi}           & 1 & MIT \\
\texttt{rusqlite}           & \href{https://github.com/rusqlite/rusqlite}{github.com/rusqlite/rusqlite}                     & 8 & MIT \\
\texttt{safe-transmute}     & \href{https://github.com/nabijaczleweli/safe-transmute-rs}{github.com/nabijaczleweli/safe-transmute-rs} & 1 & MIT \\
\texttt{scanner}            & \href{https://github.com/CasualX/scanner-rs}{github.com/CasualX/scanner-rs}                   & 1 & MIT \\
\texttt{simple-slab}        & \href{https://github.com/nathansizemore/simple-slab}{github.com/nathansizemore/simple-slab}   & 2 & MPL-2.0 \\
\texttt{sized-chunks}       & \href{https://github.com/bodil/sized-chunks}{github.com/bodil/sized-chunks}                   & 6 & MPL-2.0+ \\
\texttt{smallvec}           & \href{https://github.com/servo/rust-smallvec}{github.com/servo/rust-smallvec}                 & 2 & MIT OR Apache-2.0 \\
\texttt{vmm-sys-util}       & \href{https://github.com/rust-vmm/vmm-sys-util}{github.com/rust-vmm/vmm-sys-util}             & 1 & BSD-3-Clause \\
\texttt{xmas-elf}           & \href{https://github.com/nrc/xmas-elf}{github.com/nrc/xmas-elf}                               & 1 & Apache-2.0 OR MIT \\
\midrule
\textbf{Total}              &                                                                                               & \textbf{42} & \\
\bottomrule
\end{tabular}
\end{table}

\clearpage
\subsection{Project Statistics}
\label{sec:project_statistics}

Table~\ref{tab:project_statistics} reports project size and popularity.
Rust LOC is the maximum total over Rust source files in a crate-level vanilla
variant. Cumulative downloads were retrieved from the
\href{https://crates.io}{crates.io} API on 4 October 2026.

\begin{table}[H]
\centering
\caption{Size and popularity of the 25 source crates.}
\label{tab:project_statistics}
\small
\begin{tabular}{@{}lrr@{}}
\toprule
\textbf{Crate} & \textbf{Rust LOC} & \textbf{Downloads} \\
\midrule
\texttt{alg\_ds} & 894 & 15,080 \\
\texttt{arc-swap} & 3,971 & 346,103,784 \\
\texttt{arr} & 348 & 14,732 \\
\texttt{array-init-cursor} & 191 & 17,770,771 \\
\texttt{bitvec} & 20,679 & 293,057,926 \\
\texttt{chttp} & 1,264 & 106,084 \\
\texttt{claxon} & 3,591 & 5,043,880 \\
\texttt{fast-float} & 2,159 & 11,112,752 \\
\texttt{hashconsing} & 1,187 & 189,990 \\
\texttt{im} & 16,963 & 33,082,506 \\
\texttt{image} & 26,304 & 203,499,138 \\
\texttt{magnetic} & 1,682 & 29,535 \\
\texttt{mp3-metadata} & 1,630 & 143,471 \\
\texttt{once\_cell} & 1,331 & 1,323,917,484 \\
\texttt{ordnung} & 2,219 & 5,117 \\
\texttt{pared} & 2,471 & 12,237 \\
\texttt{redox\_uefi\_std} & 665 & 41,049 \\
\texttt{rusqlite} & 14,405 & 117,468,289 \\
\texttt{safe-transmute} & 1,379 & 5,297,974 \\
\texttt{scanner} & 809 & 3,779 \\
\texttt{simple-slab} & 210 & 20,063 \\
\texttt{sized-chunks} & 5,631 & 48,980,303 \\
\texttt{smallvec} & 2,923 & 1,256,704,458 \\
\texttt{vmm-sys-util} & 7,337 & 12,419,868 \\
\texttt{xmas-elf} & 2,212 & 3,063,690 \\
\bottomrule
\end{tabular}
\end{table}

\end{appendices}
\flushbottom


\newpage
\section*{NeurIPS Paper Checklist}

\begin{enumerate}

\Needspace{8\baselineskip}
\item {\bf Claims}
    \item[] Question: Do the main claims made in the abstract and introduction accurately reflect the paper's contributions and scope?
    \item[] Answer: \answerYes{}
    \item[] Justification: Our claims in the abstract and introduction accurately reflect the paper's contributions and scope.
    \item[] Guidelines:
    \begin{itemize}
        \item The answer \answerNA{} means that the abstract and introduction do not include the claims made in the paper.
        \item The abstract and/or introduction should clearly state the claims made, including the contributions made in the paper and important assumptions and limitations. A \answerNo{} or \answerNA{} answer to this question will not be perceived well by the reviewers. 
        \item The claims made should match theoretical and experimental results, and reflect how much the results can be expected to generalize to other settings. 
        \item It is fine to include aspirational goals as motivation as long as it is clear that these goals are not attained by the paper. 
    \end{itemize}

\Needspace{8\baselineskip}
\item {\bf Limitations}
    \item[] Question: Does the paper discuss the limitations of the work performed by the authors?
    \item[] Answer: \answerYes{}
    \item[] Justification: 
    We provide limitations of our work in 
    \autoref{sec:limitations}.
    \item[] Guidelines:
    \begin{itemize}
        \item The answer \answerNA{} means that the paper has no limitation while the answer \answerNo{} means that the paper has limitations, but those are not discussed in the paper. 
        \item The authors are encouraged to create a separate ``Limitations'' section in their paper.
        \item The paper should point out any strong assumptions and how robust the results are to violations of these assumptions (e.g., independence assumptions, noiseless settings, model well-specification, asymptotic approximations only holding locally). The authors should reflect on how these assumptions might be violated in practice and what the implications would be.
        \item The authors should reflect on the scope of the claims made, e.g., if the approach was only tested on a few datasets or with a few runs. In general, empirical results often depend on implicit assumptions, which should be articulated.
        \item The authors should reflect on the factors that influence the performance of the approach. For example, a facial recognition algorithm may perform poorly when image resolution is low or images are taken in low lighting. Or a speech-to-text system might not be used reliably to provide closed captions for online lectures because it fails to handle technical jargon.
        \item The authors should discuss the computational efficiency of the proposed algorithms and how they scale with dataset size.
        \item If applicable, the authors should discuss possible limitations of their approach to address problems of privacy and fairness.
        \item While the authors might fear that complete honesty about limitations might be used by reviewers as grounds for rejection, a worse outcome might be that reviewers discover limitations that aren't acknowledged in the paper. The authors should use their best judgment and recognize that individual actions in favor of transparency play an important role in developing norms that preserve the integrity of the community. Reviewers will be specifically instructed to not penalize honesty concerning limitations.
    \end{itemize}

\Needspace{8\baselineskip}
\item {\bf Theory assumptions and proofs}
    \item[] Question: For each theoretical result, does the paper provide the full set of assumptions and a complete (and correct) proof?
    \item[] Answer: \answerNA{}.
    \item[] Justification: Our paper does not include theoretical results.
    \item[] Guidelines:
    \begin{itemize}
        \item The answer \answerNA{} means that the paper does not include theoretical results. 
        \item All the theorems, formulas, and proofs in the paper should be numbered and cross-referenced.
        \item All assumptions should be clearly stated or referenced in the statement of any theorems.
        \item The proofs can either appear in the main paper or the supplemental material, but if they appear in the supplemental material, the authors are encouraged to provide a short proof sketch to provide intuition. 
        \item Inversely, any informal proof provided in the core of the paper should be complemented by formal proofs provided in appendix or supplemental material.
        \item Theorems and Lemmas that the proof relies upon should be properly referenced. 
    \end{itemize}

    \item {\bf Experimental result reproducibility}
    \item[] Question: Does the paper fully disclose all the information needed to reproduce the main experimental results of the paper to the extent that it affects the main claims and/or conclusions of the paper (regardless of whether the code and data are provided or not)?
    \item[] Answer: \answerYes{}
    \item[] Justification: 
    In the paper, we provide information to reproduce the main experiments: we specify the experimental setup in \autoref{sec:eval_setup}, LLM prompts in \autoref{sec:evaluation_prompts}.
    To aid in reproducibility, we also provide our framework code and dataset (\autoref{sec:artifact_availability}), and provide model settings and analysis-code locations (\autoref{sec:artifact_reproducibility}).
    \item[] Guidelines:
    \begin{itemize}
        \item The answer \answerNA{} means that the paper does not include experiments.
        \item If the paper includes experiments, a \answerNo{} answer to this question will not be perceived well by the reviewers: Making the paper reproducible is important, regardless of whether the code and data are provided or not.
        \item If the contribution is a dataset and\slash or model, the authors should describe the steps taken to make their results reproducible or verifiable. 
        \item Depending on the contribution, reproducibility can be accomplished in various ways. For example, if the contribution is a novel architecture, describing the architecture fully might suffice, or if the contribution is a specific model and empirical evaluation, it may be necessary to either make it possible for others to replicate the model with the same dataset, or provide access to the model. In general. releasing code and data is often one good way to accomplish this, but reproducibility can also be provided via detailed instructions for how to replicate the results, access to a hosted model (e.g., in the case of a large language model), releasing of a model checkpoint, or other means that are appropriate to the research performed.
        \item While NeurIPS does not require releasing code, the conference does require all submissions to provide some reasonable avenue for reproducibility, which may depend on the nature of the contribution. For example
        \begin{enumerate}
            \item If the contribution is primarily a new algorithm, the paper should make it clear how to reproduce that algorithm.
            \item If the contribution is primarily a new model architecture, the paper should describe the architecture clearly and fully.
            \item If the contribution is a new model (e.g., a large language model), then there should either be a way to access this model for reproducing the results or a way to reproduce the model (e.g., with an open-source dataset or instructions for how to construct the dataset).
            \item We recognize that reproducibility may be tricky in some cases, in which case authors are welcome to describe the particular way they provide for reproducibility. In the case of closed-source models, it may be that access to the model is limited in some way (e.g., to registered users), but it should be possible for other researchers to have some path to reproducing or verifying the results.
        \end{enumerate}
    \end{itemize}

\Needspace{8\baselineskip}
\item {\bf Open access to data and code}
    \item[] Question: Does the paper provide open access to the data and code, with sufficient instructions to faithfully reproduce the main experimental results, as described in supplemental material?
    \item[] Answer: \answerYes{}
    \item[] Justification: 
    We provide our framework code (\url{https://github.com/sfu-rsl/rust-mizan}) and our data (\url{https://huggingface.co/datasets/sfu-rsl/mizan-vanilla}).
    We also provide READMEs that describe how to run the RustMizan framework and reproduce the experimental results.
    For the data, we only provide the vanilla variants, with scripts in our framework to generate the mutated variants.
    \item[] Guidelines:
    \begin{itemize}
        \item The answer \answerNA{} means that paper does not include experiments requiring code.
        \item Please see the NeurIPS code and data submission guidelines (\url{https://neurips.cc/public/guides/CodeSubmissionPolicy}) for more details.
        \item While we encourage the release of code and data, we understand that this might not be possible, so \answerNo{} is an acceptable answer. Papers cannot be rejected simply for not including code, unless this is central to the contribution (e.g., for a new open-source benchmark).
        \item The instructions should contain the exact command and environment needed to run to reproduce the results. See the NeurIPS code and data submission guidelines (\url{https://neurips.cc/public/guides/CodeSubmissionPolicy}) for more details.
        \item The authors should provide instructions on data access and preparation, including how to access the raw data, preprocessed data, intermediate data, and generated data, etc.
        \item The authors should provide scripts to reproduce all experimental results for the new proposed method and baselines. If only a subset of experiments are reproducible, they should state which ones are omitted from the script and why.
        \item At submission time, to preserve anonymity, the authors should release anonymized versions (if applicable).
        \item Providing as much information as possible in supplemental material (appended to the paper) is recommended, but including URLs to data and code is permitted.
    \end{itemize}

\Needspace{8\baselineskip}
\item {\bf Experimental setting/details}
    \item[] Question: Does the paper specify all the training and test details (e.g., data splits, hyperparameters, how they were chosen, type of optimizer) necessary to understand the results?
    \item[] Answer: \answerYes{}
    \item[] Justification: 
    We describe the experimental setup in \autoref{sec:eval_setup}, and provide definition of the evaluation metrics in \autoref{sec:evaluation_metrics}.
    We specify the versions of the LLMs tested in \autoref{sec:eval_setup}, the prompts we used in \autoref{sec:evaluation_prompts}, and fine-tuning experimental settings in \autoref{sec:data_decontamination}.
    \item[] Guidelines:
    \begin{itemize}
        \item The answer \answerNA{} means that the paper does not include experiments.
        \item The experimental setting should be presented in the core of the paper to a level of detail that is necessary to appreciate the results and make sense of them.
        \item The full details can be provided either with the code, in appendix, or as supplemental material.
    \end{itemize}

\Needspace{8\baselineskip}
\item {\bf Experiment statistical significance}
    \item[] Question: Does the paper report error bars suitably and correctly defined or other appropriate information about the statistical significance of the experiments?
    \item[] Answer: \answerYes{}
    \item[] Justification: Appendix~\ref{sec:uncertainty_estimates} reports 95\% vulnerability-clustered bootstrap intervals for Tables~\ref{tab:main_results} and~\ref{tab:mutation_results}, including paired mutation changes. These intervals capture variation across CVEs/advisories conditional on the observed rollouts; run-to-run variability remains unmeasured.
    \item[] Guidelines:
    \begin{itemize}
        \item The answer \answerNA{} means that the paper does not include experiments.
        \item The authors should answer \answerYes{} if the results are accompanied by error bars, confidence intervals, or statistical significance tests, at least for the experiments that support the main claims of the paper.
        \item The factors of variability that the error bars are capturing should be clearly stated (for example, train/test split, initialization, random drawing of some parameter, or overall run with given experimental conditions).
        \item The method for calculating the error bars should be explained (closed form formula, call to a library function, bootstrap, etc.)
        \item The assumptions made should be given (e.g., Normally distributed errors).
        \item It should be clear whether the error bar is the standard deviation or the standard error of the mean.
        \item It is OK to report 1-sigma error bars, but one should state it. The authors should preferably report a 2-sigma error bar than state that they have a 96\% CI, if the hypothesis of Normality of errors is not verified.
        \item For asymmetric distributions, the authors should be careful not to show in tables or figures symmetric error bars that would yield results that are out of range (e.g., negative error rates).
        \item If error bars are reported in tables or plots, the authors should explain in the text how they were calculated and reference the corresponding figures or tables in the text.
    \end{itemize}

\Needspace{8\baselineskip}
\item {\bf Experiments compute resources}
    \item[] Question: For each experiment, does the paper provide sufficient information on the computer resources (type of compute workers, memory, time of execution) needed to reproduce the experiments?
    \item[] Answer: \answerYes{}
    \item[] Justification: We provide the number of tokens used for prompting proprietary LLMs in \autoref{sec:eval_setup}, and the compute used for fine-tuning in \autoref{sec:data_decontamination}.
    \item[] Guidelines:
    \begin{itemize}
        \item The answer \answerNA{} means that the paper does not include experiments.
        \item The paper should indicate the type of compute workers CPU or GPU, internal cluster, or cloud provider, including relevant memory and storage.
        \item The paper should provide the amount of compute required for each of the individual experimental runs as well as estimate the total compute. 
        \item The paper should disclose whether the full research project required more compute than the experiments reported in the paper (e.g., preliminary or failed experiments that didn't make it into the paper). 
    \end{itemize}
    
\Needspace{8\baselineskip}
\item {\bf Code of ethics}
    \item[] Question: Does the research conducted in the paper conform, in every respect, with the NeurIPS Code of Ethics \url{https://neurips.cc/public/EthicsGuidelines}?
    \item[] Answer: \answerYes{}
    \item[] Justification: We have reviewed the NeurIPS Code of Ethics, and the research conducted in the paper conform, in every respect, with the code.
    \item[] Guidelines:
    \begin{itemize}
        \item The answer \answerNA{} means that the authors have not reviewed the NeurIPS Code of Ethics.
        \item If the authors answer \answerNo, they should explain the special circumstances that require a deviation from the Code of Ethics.
        \item The authors should make sure to preserve anonymity (e.g., if there is a special consideration due to laws or regulations in their jurisdiction).
    \end{itemize}

\Needspace{8\baselineskip}
\item {\bf Broader impacts}
    \item[] Question: Does the paper discuss both potential positive societal impacts and negative societal impacts of the work performed?
    \item[] Answer: \answerYes{}
    \item[] Justification: We describe the broader impact of our work in the conclusion (\autoref{sec:conclusion}) and the risk of future contamination from public mutants in \autoref{sec:limitations}.
    \item[] Guidelines:
    \begin{itemize}
        \item The answer \answerNA{} means that there is no societal impact of the work performed.
        \item If the authors answer \answerNA{} or \answerNo, they should explain why their work has no societal impact or why the paper does not address societal impact.
        \item Examples of negative societal impacts include potential malicious or unintended uses (e.g., disinformation, generating fake profiles, surveillance), fairness considerations (e.g., deployment of technologies that could make decisions that unfairly impact specific groups), privacy considerations, and security considerations.
        \item The conference expects that many papers will be foundational research and not tied to particular applications, let alone deployments. However, if there is a direct path to any negative applications, the authors should point it out. For example, it is legitimate to point out that an improvement in the quality of generative models could be used to generate Deepfakes for disinformation. On the other hand, it is not needed to point out that a generic algorithm for optimizing neural networks could enable people to train models that generate Deepfakes faster.
        \item The authors should consider possible harms that could arise when the technology is being used as intended and functioning correctly, harms that could arise when the technology is being used as intended but gives incorrect results, and harms following from (intentional or unintentional) misuse of the technology.
        \item If there are negative societal impacts, the authors could also discuss possible mitigation strategies (e.g., gated release of models, providing defenses in addition to attacks, mechanisms for monitoring misuse, mechanisms to monitor how a system learns from feedback over time, improving the efficiency and accessibility of ML).
    \end{itemize}
    
\Needspace{8\baselineskip}
\item {\bf Safeguards}
    \item[] Question: Does the paper describe safeguards that have been put in place for responsible release of data or models that have a high risk for misuse (e.g., pre-trained language models, image generators, or scraped datasets)?
    \item[] Answer: \answerYes{}
    \item[] Justification: To avoid LLMs pretraining on mutated variants, we only release the vanilla version of our datasets.  The mutated versions should be generated using the provided scripts in the RustMizan framework.  We discuss this in \autoref{sec:artifact_card}.
    \item[] Guidelines:
    \begin{itemize}
        \item The answer \answerNA{} means that the paper poses no such risks.
        \item Released models that have a high risk for misuse or dual-use should be released with necessary safeguards to allow for controlled use of the model, for example by requiring that users adhere to usage guidelines or restrictions to access the model or implementing safety filters. 
        \item Datasets that have been scraped from the Internet could pose safety risks. The authors should describe how they avoided releasing unsafe images.
        \item We recognize that providing effective safeguards is challenging, and many papers do not require this, but we encourage authors to take this into account and make a best faith effort.
    \end{itemize}

\Needspace{8\baselineskip}
\item {\bf Licenses for existing assets}
    \item[] Question: Are the creators or original owners of assets (e.g., code, data, models), used in the paper, properly credited and are the license and terms of use explicitly mentioned and properly respected?
    \item[] Answer: \answerYes{}
    \item[] Justification: We summarize the license for the source code that compose our dataset in \autoref{sec:data_licenses}.
    \item[] Guidelines:
    \begin{itemize}
        \item The answer \answerNA{} means that the paper does not use existing assets.
        \item The authors should cite the original paper that produced the code package or dataset.
        \item The authors should state which version of the asset is used and, if possible, include a URL.
        \item The name of the license (e.g., CC-BY 4.0) should be included for each asset.
        \item For scraped data from a particular source (e.g., website), the copyright and terms of service of that source should be provided.
        \item If assets are released, the license, copyright information, and terms of use in the package should be provided. For popular datasets, \url{paperswithcode.com/datasets} has curated licenses for some datasets. Their licensing guide can help determine the license of a dataset.
        \item For existing datasets that are re-packaged, both the original license and the license of the derived asset (if it has changed) should be provided.
        \item If this information is not available online, the authors are encouraged to reach out to the asset's creators.
    \end{itemize}

\Needspace{8\baselineskip}
\item {\bf New assets}
    \item[] Question: Are new assets introduced in the paper well documented and is the documentation provided alongside the assets?
    \item[] Answer: \answerYes{}
    \item[] Justification:  
    We provide detailed READMEs and documentation for the both our framework code and our dataset.
    Information about our code and dataset is provided in \autoref{sec:artifact_card}
    \item[] Guidelines:
    \begin{itemize}
        \item The answer \answerNA{} means that the paper does not release new assets.
        \item Researchers should communicate the details of the dataset\slash code\slash model as part of their submissions via structured templates. This includes details about training, license, limitations, etc. 
        \item The paper should discuss whether and how consent was obtained from people whose asset is used.
        \item At submission time, remember to anonymize your assets (if applicable). You can either create an anonymized URL or include an anonymized zip file.
    \end{itemize}

\Needspace{8\baselineskip}
\item {\bf Crowdsourcing and research with human subjects}
    \item[] Question: For crowdsourcing experiments and research with human subjects, does the paper include the full text of instructions given to participants and screenshots, if applicable, as well as details about compensation (if any)? 
    \item[] Answer: \answerNA{}
    \item[] Justification: Our paper does not involve crowdsourcing nor research with human subjects.
    \item[] Guidelines:
    \begin{itemize}
        \item The answer \answerNA{} means that the paper does not involve crowdsourcing nor research with human subjects.
        \item Including this information in the supplemental material is fine, but if the main contribution of the paper involves human subjects, then as much detail as possible should be included in the main paper. 
        \item According to the NeurIPS Code of Ethics, workers involved in data collection, curation, or other labor should be paid at least the minimum wage in the country of the data collector. 
    \end{itemize}

\Needspace{8\baselineskip}
\item {\bf Institutional review board (IRB) approvals or equivalent for research with human subjects}
    \item[] Question: Does the paper describe potential risks incurred by study participants, whether such risks were disclosed to the subjects, and whether Institutional Review Board (IRB) approvals (or an equivalent approval/review based on the requirements of your country or institution) were obtained?
    \item[] Answer: \answerNA{}
    \item[] Justification: Our paper does not involve crowdsourcing nor research with human subjects.
    \item[] Guidelines:
    \begin{itemize}
        \item The answer \answerNA{} means that the paper does not involve crowdsourcing nor research with human subjects.
        \item Depending on the country in which research is conducted, IRB approval (or equivalent) may be required for any human subjects research. If you obtained IRB approval, you should clearly state this in the paper. 
        \item We recognize that the procedures for this may vary significantly between institutions and locations, and we expect authors to adhere to the NeurIPS Code of Ethics and the guidelines for their institution. 
        \item For initial submissions, do not include any information that would break anonymity (if applicable), such as the institution conducting the review.
    \end{itemize}

\Needspace{8\baselineskip}
\item {\bf Declaration of LLM usage}
    \item[] Question: Does the paper describe the usage of LLMs if it is an important, original, or non-standard component of the core methods in this research? Note that if the LLM is used only for writing, editing, or formatting purposes and does \emph{not} impact the core methodology, scientific rigor, or originality of the research, declaration is not required.
    \item[] Answer: \answerYes{}
    \item[] Justification: We specify the LLMs we use for evaluate on our benchmark in \autoref{sec:eval_setup}, and the prompts we use with the LLMs in \autoref{sec:evaluation_prompts}.  For data contamination experiments, we fine-tune open-source LLMs as specified in 
    \autoref{sec:data_decontamination}.
    \item[] Guidelines:
    \begin{itemize}
        \item The answer \answerNA{} means that the core method development in this research does not involve LLMs as any important, original, or non-standard components.
        \item Please refer to our LLM policy in the NeurIPS handbook for what should or should not be described.
    \end{itemize}

\end{enumerate}

\end{document}